\documentstyle[psfig]{mn}

\def\simg{\mathrel{%
      \rlap{\raise 0.511ex \hbox{$>$}}{\lower 0.511ex \hbox{$\sim$}}}}
\def\siml{\mathrel{%
      \rlap{\raise 0.511ex \hbox{$<$}}{\lower 0.511ex \hbox{$\sim$}}}}
\def\Meszaros{M\'esz\'aros } 
\def\eq{eq$.$} \def\eqs{eqs$.$} \def\etal{et al$.$ } \def\eg{e$.$g$.$ } \def\ie{i$.$e$.$ }

\def\epsi{\epsilon_i} \def\epsB{\epsilon_B} \def\cm3{\,\rm cm^{-3}} 
\def\E{{\cal E}} \def\ER{{\cal E}^{(R)}} \def\EO{{\cal E}^{(O)}}
\def\M{{\cal M}} \def\Mi{{\cal M}_i} 

\def\nuiO{\nu_i^{(O)}} \def\nuiOtb{\nu_i^{(O)}(t_b)}
\def\nucO{\nu_c^{(O)}} \def\nucOtb{\nu_c^{(O)}(t_b)} 
\def\nuiRtr{\nu_i^{(R)}(t_r)} \def\nucRtr{\nu_c^{(R)}(t_r)}

\def\FpOtb{F_p^{(O)}(t_b)} \def\FrO{F_r^{(O)}} \def\FrR{F_r^{(R)}} 
\def\Frtr{F_r (t_r)} \def\Fotb{F_o (t_b)} 

\def\nuiRS{\nu_i^{(R)}} \def\nuiFS{\nu_i^{(F)}} 
\def\nuiFStb{\nu_i^{(F)}(t_b)} \def\nuctb{\nu_c(t_b)} 

\def\reference{\bibitem}

\onecolumn

\begin{document}

\title[The Slow Decay of Some GRB Radio Afterglows]
      {The Slow Decay of Some Radio Afterglows, a Puzzle for the Simplest GRB Fireball Model}

\author[A. Panaitescu \& P. Kumar]{A. Panaitescu and P. Kumar \\
        Department of Astronomy, University of Texas, Austin, TX 78712}

\maketitle

\begin{abstract}

 The decay of half of the GRB radio afterglows with long temporal monitoring is significantly 
  slower than at optical frequencies, contrary to what is expected in the simplest fireball model. 
 We investigate four ways to decouple the radio and optical decays:
  an evolving index of the power-law distribution of the shock-accelerated electrons, 
  the presence of a spectral break between the two domains, 
  a structured outflow,
  and a long-lived reverse shock contribution to the afterglow emission.  
 For most afterglows, the first scenario cannot accommodate all properties of the afterglow emission. 
 If the spectral break of the second scenario is the cooling frequency, then observations require 
  that, as the fireball decelerates, the parameter for the minimal electron energy decreases and the 
  magnetic field strength is constant or increases. The latter behavior seems implausible and is 
  alleviated if the circumburst density increases outward. 
 In the framework of the third scenario, the optical afterglow arises from a more energetic, narrow 
  outflow core while the radio emission comes from a more extended envelope. This scenario is at best
  marginally consistent with the general properties of the radio and optical afterglow emissions and 
  requires a total electron energy exceeding equipartition, thus it does not provide an acceptable solution.
 In the fourth scenario, it is assumed that the radio afterglow emission arises in the reverse shock
  propagating in a steady stream of ejecta, which catch up with the decelerating GRB remnant.
  This scenario can accommodate the properties of the afterglows with slow radio decays and requires 
  that the injected energy is less than or comparable to the initial fireball energy, while other 
  afterglow parameters have reasonable values. We find the reverse--forward shock scenario to be
  the most viable explanation for the shallow decay observed in some radio afterglows.
 For a jet, the transition to a semi-relativistic motion mitigates the radio decay, however this
  scenario would work only when the slower radio decay is observed well after the steeper optical
  fall-off. A structured outflow with a relativistic core, yielding a fast decaying optical emission,
  and a non-relativistic envelope, producing a slowly falling-off radio afterglow, is not a viable
  solution, as it fails to decouple the radio and optical decays.
 
\end{abstract}


\section{Introduction}

 Since their first discovery, the optical afterglows of Gamma-Ray Bursts (GRBs) have displayed 
a time evolution which is consistent with the predictions of the fireball model (\Meszaros \& 
Rees 1997). The power-law decay expected for a spherical outflow has been first observed in
GRB 970228 (Wijers, Rees \& \Meszaros 1997). A tight collimation of the GRB outflow (\ie a narrow
jet) was shown to yield a steepening ("break") of the afterglow light-curve fall-off (Rhoads 1999), 
which subsequently was observed in GRB 990123 (Kulkarni \etal 1999). To date, there are about ten
afterglows exhibiting an optical light-curve break and a smaller number of sufficiently well 
monitored afterglows whose light-curves are pure power-laws. In all cases where the optical 
spectral slope was determined, it was found to be consistent (within the fireball/jet model) 
with the optical light-curve decay rate.

 Further tests of the basic afterglow model require observations at earlier times, within the first 
couple of hours after the burst, and at other wavelengths. The number of GRB afterglows with early
observations will be greatly enhanced by the Swift satellite. On the latter avenue, the monitoring
of about ten $X$-ray afterglows has shown that their behavior is consistent with that observed in
the optical, without major surprises for the fireball model. However, it should be noted that 
$X$-ray observations rarely cover more than one decade in time. Wider time ranges have been
reached for about ten radio afterglows, half of them exhibiting a decay which is much slower 
than in the optical, which is a rather puzzling feature for the fireball model. That radio 
light-curves fall-off after about 10 days, strongly indicate that peak of the synchrotron spectrum 
is redward of the radio domain after that time. Then, within the fireball model, one expects nearly
equally fast decays in the radio and in the optical, contrary to some observations. 

 The purpose of this paper is to analyze in more detail whether the fireball/jet model in its
simplest form can explain the discrepancy between the radio and optical decays observed in some 
afterglows (\S\ref{standard}) and to explore departures from the usual assumptions that could 
accommodate such discrepancies (\S\ref{nonstandard}). The expression "simplest fireball model" 
will be used to designate an outflow $i)$ with a uniform angular distribution of the kinetic 
energy, $ii)$ without energy injection during the afterglow phase, $iii)$ interacting with a 
circumburst medium whose density changes monotonically with radius, $iv)$ with un-evolving and 
uniform parameters for the electron distribution and magnetic field behind the forward shock. 
For an easier understanding of the following equations, we list in Table 1 the definitions of the 
quantities which will be encountered more often.

\section{Radio Light-Curves in the Simplest Fireball/Jet Model}
\label{standard}

 For a spherical, uniform fireball, or a wide jet whose dynamics is not yet affected by its 
lateral spreading, the synchrotron characteristic frequency corresponding to the lowest energy 
electrons accelerated by the forward shock (hereafter called "injection" frequency) is given by 
(\eg Panaitescu \& Kumar 2000)
\begin{equation}
 \nu_i = 10^{12}\; (z+1)^{1/2} \left( \frac{\E}{10^{53}\, {\rm ergs}} \right)^{1/2} 
                         \left( \frac{\epsi}{0.03} \right)^2 
                         \left( \frac{\epsB}{10^{-3}} \right)^{1/2} 
                         \; t_d^{-3/2}\; {\rm Hz} \;
\label{nui}
\end{equation}
where $\E$ is the fireball kinetic energy per solid angle, $t_d$ is the observer frame time 
measured in days, $z$ the burst redshift, $\epsB$ is the fractional post-shock energy stored 
in the magnetic field, and $\epsi$ parameterizes the lowest energy $\gamma_i m_e c^2$ of the 
power-law distribution of shock-accelerated electrons, representing the fraction of the energy 
in the downstream fluid residing in electrons if they all had the same random Lorentz factor 
$\gamma_i$, \ie $\gamma_i m_e c^2 n_e = \epsi \times \Gamma m_p c^2 n_p$, where $n_e = n_p$
are the electron and proton number densities in the post-shock fluid.

 The synchrotron characteristic frequency corresponding to an electron energy for which the
radiative losses timescale equals the fireball age (called the "cooling" frequency) is
\begin{equation}
 \nu_c = 4 \times 10^{15}\; (z+1)^{-1/2} \left( \frac{\E}{10^{53}\, {\rm ergs}} \right)^{-1/2} 
                         \left( \frac{n}{1\cm3} \right)^{-1} 
                         \left( \frac{\epsB}{10^{-3}} \right)^{-3/2} 
                         (Y+1)^{-2}\; t_d^{-1/2}\; {\rm Hz} \;,
\label{nuc0}
\end{equation}
for a homogeneous circumburst medium of particle density $n$, and 
\begin{equation}
 \nu_c = 4 \times 10^{16}\; (z+1)^{-3/2} \left( \frac{\E}{10^{53}\, {\rm ergs}} \right)^{1/2} 
                         A_*^{-2} 
                         \left( \frac{\epsB}{10^{-3}} \right)^{-3/2} 
                         (Y+1)^{-2}\; t_d^{1/2} \; {\rm Hz} \;,
\label{nuc2}
\end{equation}
for an $r^{-2}$ wind-like density profile, $A_*$ being the ratio of the mass loss rate to 
speed of the wind, in units of $10^{-8} {\rm (M_\odot/ yr^{-1}) (km/s)^{-1}}$.
In the above equations, $Y$ is the Compton parameter. 

 If the afterglow is optically thin at the peak of the spectrum, the synchrotron self-absorption 
frequency is
\begin{equation}
 \nu_a = 8 \; (z+1)^{-1} \left( \frac{\E}{10^{53}\, {\rm ergs}} \right)^{1/5} 
                         \left( \frac{n}{1\cm3} \right)^{3/5} 
                         \left( \frac{\epsi}{0.03} \right)^{-1} 
                         \left( \frac{\epsB}{10^{-3}} \right)^{1/5} \; {\rm GHz} \;,
\label{nua0a}
\end{equation}
for a homogeneous external medium, and
\begin{equation}
 \nu_a = 3 \; (z+1)^{-2/5} \left( \frac{\E}{10^{53}\, {\rm ergs}} \right)^{-2/5} 
                         A_*^{6/5} 
                         \left( \frac{\epsi}{0.03} \right)^{-1} 
                         \left( \frac{\epsB}{10^{-3}} \right)^{1/5} 
                         \; t_d^{-3/5} \; {\rm GHz} \;,
\label{nua2a}
\end{equation}
for a wind. In the derivation of the above equations it was assumed that $\nu_a < \nu_i < \nu_c$. 
For a high magnetic field parameter $\epsB$, the cooling frequency may be below the injection, 
in which case $\nu_a$ is given by 
\begin{equation}
 \nu_a = 30 \; (z+1)^{-1/2} \left( \frac{\E}{10^{53}\, {\rm ergs}} \right)^{7/10} 
                         \left( \frac{n}{1\cm3} \right)^{11/10} 
                         \left( \frac{\epsB}{10^{-1}} \right)^{6/5} 
                         (Y+1)\; t_d^{-1/2}\; {\rm GHz} \;,
\label{nua0r}
\end{equation}
for a homogeneous external medium,
\begin{equation}
 \nu_a = 6 \; (z+1)^{3/5} \left( \frac{\E}{10^{53}\, {\rm ergs}} \right)^{-2/5} 
                         A_*^{11/5} 
                         \left( \frac{\epsB}{10^{-1}} \right)^{6/5} 
                         (Y+1)\; t_d^{-8/5} \; {\rm GHz} \;,
\label{nua2r}
\end{equation}
for a wind. At late times, the injection frequency falls below the absorption frequency.
In this case, the latter evolves as $\nu_a \propto t^{-(3p+2)/(2p+8)}$ (homogeneous medium) 
and $\nu_a \propto t^{-3(p+2)/(2p+8)}$ (wind medium), where $p$ is the exponent of the power-law 
energy distribution of the electrons accelerated at the forward shock: $dN/d\gamma \propto 
\gamma^{-p}$, $\gamma$ being the electron random Lorentz factor.
 
 The flux at the peak frequency ($= \min \{ \nu_i, \nu_c \}$) of the synchrotron spectrum is
\begin{equation}
 F_p = 30 \; (z+1)       \left( \frac{d_L(z)}{d_L(z=1)} \right)^{-2}
                         \left( \frac{\E}{10^{53}\, {\rm ergs}} \right)
                         \left( \frac{n}{1\cm3} \right)^{1/2} 
                         \left( \frac{\epsB}{10^{-3}} \right)^{1/2} \; {\rm mJy} \;,
\label{fp0}
\end{equation}
for a homogeneous external medium and
\begin{equation}
 F_p = 14 \; (z+1)^{3/2} \left[ \frac{d_L(z)}{d_L(z=1)} \right]^{-2}
                         \left( \frac{\E}{10^{53}\, {\rm ergs}} \right)^{1/2} 
                         A_*
                         \left( \frac{\epsB}{10^{-3}} \right)^{1/2} 
                         \; t_d^{-1/2} \; {\rm mJy} \;,
\label{fp2}
\end{equation}
for a wind-type medium, where $d_L(z)$ is the luminosity distance. 

 For a spreading jet, the time-dependence of the spectral breaks is given by 
\begin{equation}
 \nu_a \propto t^{-1/5}\;\; {\rm for}\;\; \nu_a < \nu_i < \nu_c\;, \quad 
 \nu_a \propto t^{-6/5}\;\; {\rm for}\;\; \nu_a < \nu_c < \nu_i \;, \quad
  \nu_i \propto t^{-2}\;\;, \quad \nu_c = const \;, 
\end{equation}
while the synchrotron peak flux evolution is $F_p \propto t^{-1}$.
For $\nu_i < \nu_a$, the self-absorption frequency evolution is $\nu_a \propto t^{-2(p+1)/(p+4)}$. 
 
 To assess which radio afterglows exhibit a decay consistent with that seen at optical frequencies,
we list in Table 2 the expected radio light-curves for spherical and collimated outflows, a 
homogeneous or wind-like medium, and for the only three likely orderings of the break frequencies 
a few days after the burst. These light-curve indices were derived from the shape of the synchrotron 
spectrum (Sari, Piran \& Narayan 1998) and the evolution of the break frequencies and peak flux. 
Most of the cases shown in Table 2 for $\nu_a < \min \{ \nu_i,\nu_c \}$ have been presented in other 
works (\eg \Meszaros, Rees \& Wijers 1998; Sari, Piran \& Narayan 1998; Rhoads 1999; Sari, Piran \& 
Halpern 1999; Chevalier \& Li 2000; Panaitescu \& Kumar 2000).

\subsection{Comparison with Observations}
\label{obs}

 As shown in Table 2, the self-absorbed, low frequency emission rises in time in most cases, as the 
optical thickness of the swept-up medium decreases, while the afterglow light-curve at higher 
frequencies, above both $\nu_a$ and $\nu_i$, falls-off. For observing frequencies higher than both 
$\nu_i$ and $\nu_c$, the light-curve is $t^{-(3p-2)/4}$ for a fireball (either type of medium) and 
$t^{-p}$ for a spreading jet. Such behaviors are observed in the radio, optical, and X-ray light-curves 
of GRB afterglows, as illustrated in Figure 1, displaying all afterglows with a good time coverage in 
the radio. A flat radio light-curve or a rise slower than $t^{1/4}$ can be noticed for most 
afterglows until 20--30 days (70 days for 970508), indicating that $\nu_a$ is never well above 
the radio domain, as in this case the radio emission should exhibit a sharper rise.

 The late time radio emissions of afterglows 970508, 000418, and 021004 exhibit the same decay as in 
the optical, as expected when both domains lie above the injection frequency and the source is optically 
thin in the radio, thus these afterglows are readily consistent with the expectations for the simplest 
fireball model. However, for all other afterglows, the radio emission falls-off significantly slower 
than in the optical. 

 If the cooling frequency $\nu_c$ lies above radio ($\nu_r$) but below optical ($\nu_o$), then 
the difference between the decay slopes of the optical ($t^{-\alpha_o}$) and radio ($t^{-\alpha_r}$) 
light-curves is $\Delta \alpha_{or} = \alpha_o - \alpha_r = 0.25$ for a homogeneous medium and 
$\Delta \alpha_{or} = - 0.25$ for a wind, the former value being close to that observed in 980703.
\footnote{
 If electron cooling is dominated by inverse Compton scatterings and electrons are in the slow
 cooling regime, then $\nu_c$ decreases slower for a homogeneous medium and increases faster 
 for a wind than given in equations (\ref{nuc0}) and (\ref{nuc2}) (Panaitescu \& Kumar 2000), 
 which mitigates the fall-off of the optical light-curve and yields a smaller $\Delta \alpha_{or}$, 
 thus this case does not accommodate the observations better} .
This possibility can be tested by noting that $\nu_c < \nu_i$ implies that the optical and 
X-ray decays have the same slopes. The upper limit at 5 days on the X-ray emission of 980703 
indicates an X-ray slope larger than $\alpha_x = 1.22$ (Figure 1), thus it may be consistent 
with the optical slope $\alpha_o = 1.37$. 

 Index differences larger than $\Delta \alpha_{or} = 1/4$, such as those seen in the afterglows 
991208, 991216, 000301, 000926, and 010222, require the presence of a stronger spectral break 
between radio and optical. However, to obtain a falling-off radio light-curve, the source must 
be optically thin at radio frequencies (Table 2), therefore that spectral break may only be 
$\nu_i$. The $\nu_a < \nu_r < \nu_i$ cases that yield a decaying radio emission are: 
$i)$ a spreading jet with $\nu_a < \nu_r < \nu_i$ and $\nu_c < \nu_i$,
$ii)$ a fireball interacting with a wind and $\nu_a < \nu_r < \nu_c < \nu_i$, 
$iii)$ a spreading jet with $\nu_a < \nu_r < \nu_i < \nu_c$. 
Below, we analyze each case in more detail.

\subsection{Spreading jet, $\nu_a < \nu_r < \nu_i < \nu_o$ and $\nu_c < \nu_i$} 
\label{jet1}

 In this case the radio decay slope is $\alpha_r = 1$, similar to that exhibited by the afterglow
991208 and 000301. Prior to the time $t_b$ when the jet edge becomes visible (the jet lateral spreading 
also becomes important at this time if the jet has sharp edges), $\nu_i$ is given by equation (\ref{nui}), 
while for $t > t_b$,  $\nu_i \propto t^{-2}$. Therefore
\begin{equation}
 \nu_i (t > t_b)= 10\; (z+1)^{1/2} \left( \frac{\E}{10^{53}\, {\rm ergs}} \right)^{1/2} 
                         \left( \frac{\epsi}{0.1} \right)^2 
                         \left( \frac{\epsB}{0.1} \right)^{1/2}  \;
                         \left( \frac{t_b}{1\, d} \right)^{1/2}\; 
                         \left( \frac{t}{100\, d} \right)^{-2}\; {\rm GHz} \;
\label{nui100}
\end{equation}
implying that, for $t_b \sim 1$ day (as observed), the electron and magnetic field parameters must 
be close to equipartition, $\epsi = \epsB = 0.1$, to maintain $\nu_i \simg 10$ GHz until the last 
radio measurement. Because $\nu_c$ is constant for a spreading jet, the condition $\nu_c (100 d) 
< \nu_i (100 d)$ is equivalent with $\nu_c (t_b) < \nu_i (t_b) (t_b/100 d)^2$. Using equations 
(\ref{nuc0}) and (\ref{nuc2}) with $Y \siml 1$ (implied by $\epsi \sim \epsB$), it leads to 
$n \simg 200\, t_{j,d}^{-1} \cm3$ for a homogeneous medium or $A_* \simg 30$ for a wind. Then, 
equations (\ref{nua0r}) and (\ref{nua2r}) give $\nu_a (t_b) \simg 10^{13}\; t_{j,d}^{-8/5}$ Hz. 
At $t > t_b$, $\nu_a \propto t^{-6/5}$, therefore $\nu_a (100 d) \simg 50\, t_{j,d}^{-2/5}$ GHz, 
\ie the source is optically thick at 8 GHz at all observing times. Thus the conditions $\nu_a < \nu_r$ 
and $\nu_c < \nu_i$ for $t \siml 100$ days are incompatible.

 Furthermore, the condition $\nu_c (100 d) < \nu_i (100 d)$ and equation (\ref{nui100}) imply that 
$\nu_c (t_b) \siml 10$ GHz. First, assume that $\nu_c (t_b) \sim 10$ GHz. Then the radio flux $F_r$ 
at $t_b$ is the synchrotron peak flux given in equations (\ref{fp0}) and (\ref{fp2}). The above 
values of $\epsB$, $n$, and $A_*$ yield $F_r (t_b) \simg 10\; t_{j,d}^{-1/2}$ Jy, decaying as 
$t^{-1}$ at $t > t_b$ to $F_r (100 d) \simg 100\; t_{j,d}^{1/2}$ mJy. 
For $\nu_a (100 d) \simg 50$ GHz, self-absorption reduces the 8 GHz flux to a few mJy, well above 
the 0.1 mJy radio flux observed at 100 days. The inconsistency between the expected and observed 
radio fluxes is not alleviated if $\nu_c (t_b) < 10$ GHz (as assumed above) because this requires 
a higher $\E$ or a denser external medium ($\epsB$ is already maximal), which, due to that
$F_\nu (t_b) \propto \E^{3/4}$ at $\nu_c  < \nu < \nu_i$, can only increase the radio flux.

 Thus we reach the conclusion that a spreading jet with  $\nu_a < \nu_r < \nu_i < \nu_o$ and 
$\nu_c < \nu_i$ cannot accommodate the radio and optical decays seen in 991208 and 000301. 
We note that, if the optical emission seen in 991208 after 10 days is not from the afterglow 
but to a supernova contribution (Castro-Tirado \etal 2001), then the radio and optical fall-offs 
are not measured simultaneously, leaving the possibility of a mildly relativistic jet during the 
radio decay, which yields at $\nu_r > \nu_i$ a decay slower than the $t^{-p}$ expected for
very relativistic jets. However, $p \siml 1.5$ would be required in this case, which is clearly 
inconsistent with the steep $t^{-2.5}$ optical decay of 991208 observed until 10 days.

\subsection{Fireball in wind-like medium, $\nu_a < \nu_r < \nu_c < \nu_i < \nu_o$}
\label{fballwind}

 In this case $\alpha_r = 2/3$, consistent with the radio decay seen for 991216 and 000926 until
100 days. Although the decrease of $\nu_i$ for a fireball ($t^{-3/2}$) is slower than for a spreading
jet ($t^{-2}$), electron and magnetic field parameters close to equipartition are still required in 
this case to ensure that $\nu_c < \nu_i$ until 100 days. The reason is that, for a fireball interacting 
with a wind, $\nu_c$ increases as $t^{1/2}$, thus the $\nu_i/\nu_c$ ratio decreases equally fast for 
either a fireball in a wind or for a spreading jet. Consequently, this model requires the same dense 
winds, with $A_* \simg 30$, leading to $\nu_a = 10\; (t/100d)^{-8/5}$ GHz (\eq [\ref{nua2r}]), \ie a 
source that is not optically thin at 8 GHz, and to $F_p \sim 1\; (t/100d)^{-1/2}$ Jy (\eq [\ref{fp2}]), 
well above the observed radio fluxes. Thus the fireball model for the radio decay of 991216 and 000926 
encounters the same problems as the jet model for 991208 and 000301.

\subsection{Spreading jet, $\nu_a < \nu_r < \nu_i < \nu_c < \nu_o$}
\label{jet2}

 The radio decay slope expected for this case, $\alpha_r = 1/3$, is close to that observed in 010222 
at 5--100 days. As for the first model above, the condition $\nu_i (100 d) > 10$ GHz for a spreading 
jet with $t_b \siml 1$ day requires that $\epsi$ and $\epsB$ are close to equipartition (\eq [\ref{nui100}]).

 The other two conditions involved in this hypothesis -- $\nu_a (t_b) < 10$ GHz and $\nu_i (t_b) 
< \nu_c (t_b)$ -- set upper bounds on the external density $n$ and wind parameter $A_*$.
From equations  (\ref{nuc0})--(\ref{nua2a}) we find the former 
condition being more restrictive, leading to $n \siml 10 \cm3$ for a homogeneous medium and 
$A_* \siml 4\; t_{j,d}^{1/2}$ for a wind. Lower bounds on the same parameters are set by that 
the peak flux at 100 days should exceed that observed at 8 GHz for 010222 (\ie 0.03 mJy). 
From equations (\ref{fp0}), (\ref{fp2}), and the $F_p \propto t^{-1}$ expected for a spreading 
jet, we find that $n \simg 10^{-4}\; t_{j,d}^{-2} \cm3$ for a homogeneous medium or $A_* \simg 0.02\; 
t_{j,d}^{-1/2}$ for a wind. 
Given that $\nu_i (100d) \simg 10$ GHz, $n$ and $A_*$ should be close to these lower limits, 
implying thus a  very tenuous external medium.

 Substituting these lower limits in equations (\ref{nuc0}) and (\ref{nuc2}), it can be shown that 
$\nu_c (t_b)\simg 5\times 10^{16}\; t_{j,d}^{3/2}$ Hz for either type of medium, \ie the cooling 
frequency is between the optical and X-ray domains. As the cooling frequency is constant for a 
spreading jet, the optical and X-ray decay slopes are expected to be equal. This is consistent 
with the observations of 010222 at 0.4--3 days (Figure 1). 

 Thus the radio, optical, and X-ray decay slopes of 010222 may be accommodated by a spreading 
jet with $\nu_a < \nu_r < \nu_i < \nu_c < \nu_o < \nu_x$. We note that numerical calculations of 
light-curves from spreading jets show that, when the dynamics is accurately calculated and the 
analytical approximation of constant jet radius (Sari, Piran \& Halpern 1999) is relaxed, the jet 
light-curve at $\nu < \nu_i$ is flat instead of decaying as $t^{-1/3}$, thus it may be that, in 
fact, the radio fall-off of 010222 is not consistent with a spreading jet.

\section{Beyond the Simplest Fireball Model}
\label{nonstandard}

 As argued in the previous section, the radio and optical decay observed for the afterglows 991208,
991216, 000301, 000926, and 010222 appear incompatible with the simplest fireball model. For these
afterglows, at least one of the assumptions usually made in the calculation of analytical 
light-curves must be relaxed. In scenarios where the radio an optical emissions arise from the same 
population of electrons, the ratio of the afterglow emission at these frequencies depends on the slope 
of the afterglow spectrum and on the evolution of all spectral break frequencies located between the
radio and optical domain. In scenarios where the radio and optical emissions arise in different regions 
of the GRB outflow, the segregation may be either angular or radial, \ie in the directions perpendicular 
and parallel, respectively, to that of the outflow motion. 
 
 Thus, we consider four ways in which the difference $\alpha_o-\alpha_r$ between the radio and optical 
decay indices may depart from expectations:
\begin{enumerate}
\item radio and optical arise from the same electron population.
  \begin{enumerate} 
  \item there is no spectral break between the radio and the optical domains. Then the variation 
        in time of the slope of the synchrotron spectrum, which is determined by that 
        of the electron distribution $dN/d\gamma \propto \gamma^{-p}$,
  \item there is a break between radio and optical, whose evolution is different than that 
        expected in the simplest afterglow model (\S\ref{standard}),
  \end{enumerate}
\item radio and optical arise from different electron populations.
  \begin{enumerate} 
  \item the anisotropic distribution of the energy per solid angle $\E$ across the outflow structure, 
      so that the radio and optical emissions arise from different parts of the outflow, moving at 
      different Lorentz factors (for the same photon-arrival time),
   \item the existence of a long-lived flow of slower ejecta which are energized by the reverse shock 
      and dominate the radio afterglow emission, while the optical afterglow arises in the forward 
      shock propagating into the circumburst medium.
  \end{enumerate}
\end{enumerate}
We analyze below these possibilities for the five afterglows (991208, 991216, 000301, 000926,
and 010222) for which the radio and optical decay indices are incompatible with each other in the 
simplest fireball model. We also discuss the effect of non-relativistic motion on the decoupling
of the radio and optical light-curve decays.

\subsection{Electron Distribution with Time-Varying Slope}
\label{varp}

 In a scenario employing an evolving $p$, the steepening of the optical light-curve decay observed 
in the afterglows 991216, 000301, 000926, and 010222 could arise from a non-monotonic evolution of 
$p$ and not necessarily from a collimated outflow. We consider first a spherical outflow and discuss 
later the case of a jet.  

 If the onset of the radio fall-off at $t_r \sim 10$ days is identified with the passage of $\nu_i$ 
(which is the most natural explanation) then, for a spherical outflow, $\nu_i (t) = \nu_r 
(t/t_r)^{-3/2}$, independent of the type of external medium. 
For a power-law density profile of the circumburst medium,
\begin{equation}
  n (r) \propto r^{-s}\, \quad  s < 3 \;,
\label{nr}
\end{equation}
the evolutions of the peak flux $F_p$ and cooling frequency are $F_p \propto t^{-s/(8-2s)}$ and 
$\nu_c \propto t^{-(4-3s)/(8-2s)}$, respectively. Substituting in the afterglow flux at $\nu > \nu_a$:
\begin{equation}
 F_\nu = F_p \times \left\{ \begin{array}{ll} 
                   (\nu/\nu_i)^{-(p-1)/2} & \nu_i < \nu < \nu_c \\
                   (\nu_c/\nu_i)^{-(p-1)/2} (\nu/\nu_c)^{-p/2} & \nu_c < \nu
                     \end{array} \right. \;,
\label{spek}
\end{equation}
they lead to a light-curve decay index at frequency $\nu$ given by
\begin{equation}
 \alpha_\nu (t) \equiv - \frac{d\ln F_\nu}{d\ln t} = \frac{3}{4}p(t) +
     \left(\frac{1}{2} \ln \frac{\nu}{\nu_r} + \frac{3}{4} \ln \frac{t}{t_r} \right) \frac{dp}{d\ln t} 
       - \frac{12-5s}{4(4-s)} \quad \left\{ + \frac{4-3s}{4(4-s)} \right\} \;,
\label{alfa}
\end{equation}
where the additional last term in the $rhs$ applies only for $\nu_c < \nu$. 

 Evidently, an evolving index $p$ can yield a pure power-law afterglow light-curve in only one
observational domain. Given that optical measurements cover the widest time range and usually
have the smallest uncertainties, we determine the evolution of $p$ by requiring that the optical
light-curve is a power-law, as observed, \ie that the optical decay index given in equation
(\ref{alfa}) does not vary in time. Equation (\ref{alfa}) can be recast as
\begin{equation}
  \frac{dp}{p_*-p} = \frac{d\ln t}{\ln \frac{t}{t_r} + \frac{2}{3} \ln \frac{\nu_o}{\nu_r}} \;, 
     \quad {\rm where} \quad
  p_* \equiv \frac{4}{3}\alpha_o +  \frac{12-5s}{3(4-s)} \quad  \left\{ - \frac{4-3s}{3(4-s)} \right\} 
\label{dpdt}
\end{equation}
is the electron index that would be required by the observed $\alpha_o$ if $p$ were constant.
Integrating the above equation yields 
\begin{equation}
  p(t) = p_* - (p_*-p_o) \frac{ \ln \frac{t_o}{t_r} + \frac{2}{3} \ln \frac{\nu_o}{\nu_r} }
                              { \ln \frac{t}{t_r}   + \frac{2}{3} \ln \frac{\nu_o}{\nu_r} } \;,
\label{pt}
\end{equation}
where $p_o$ is the value of $p$ at some time $t_o$ when the optical slope $\beta_o$ was measured:
$p_o = 2\beta_o + 1$ if $\nu_o < \nu_c$ and $p_o = 2\beta_o$ if $\nu_c < \nu_o$. 

 There are a few consequences stemming from equations (\ref{alfa}), (\ref{dpdt}), and (\ref{pt}). 
For $p < p_*$, the index $p$ increases in time, while for $p > p_*$ it decreases, in either case 
approaching asymptotically the value $p_*$. The light-curve index at an arbitrary frequency $\nu$ 
satisfies
\begin{equation}
  \alpha_\nu = \alpha_o + \frac{1}{2}\ln \frac{\nu}{\nu_o} \frac{dp}{d\ln t} \quad  
                \left\{ \pm \frac{4-3s}{4(4-s)} \right\} \;,
\label{alpha}
\end{equation}
where the sign of the last term in the $rhs$ is "+" for $\nu_o < \nu_c < \nu$, and "--" for 
$\nu < \nu_c < \nu_o$, otherwise that term is absent. Substituting equations (\ref{dpdt}) and 
(\ref{pt}) in (\ref{alpha}), we obtain
\begin{equation}
  \alpha_\nu = \alpha_o + \frac{1}{2} (p_*-p_o) \ln \frac{\nu}{\nu_o}
                          \frac{ \ln \frac{t_o}{t_r} + \frac{2}{3} \ln \frac{\nu_o}{\nu_r} }
                               { \ln \frac{t}{t_r}   + \frac{2}{3} \ln \frac{\nu_o}{\nu_r} }
                \quad        \left\{ \pm \frac{4-3s}{4(4-s)} \right\} \;.
\label{alfanu}
\end{equation}
Therefore, a radio decay significantly shallower than in the optical ($\alpha_r < \alpha_o - 1/4$) 
requires that $p < p_*$, implying an increasing $p$ and an optical spectrum that softens in time
($d \beta_o /dt > 0$). Furthermore, for $p < p_*$, equation (\ref{alfanu}) gives that the afterglow
decay in the $X$-ray should be steeper than in the optical ($\alpha_x > \alpha_o$). Also for $p < p_*$, 
the evolution of $\alpha_\nu$, $d\alpha_\nu/dt \propto (p_*-p) \ln (\nu_o/\nu)$, shows that the radio 
decay should steepen in time ($d\alpha_r/dt > 0$), while the $X$-ray decay should flatten 
($d\alpha_x/dt < 0$). 

 The above consequences of requiring that $\alpha_r < \alpha_o$ can be used to test the scenario
of an evolving index $p$. Using the light-curve characteristics given in Table 3, we find that a radio
decay consistent with the observations is obtained only if $s=0$ and $\nu_c < \nu_o$, for the
GRB afterglows
\begin{enumerate}
 \item 991208, for which equation (\ref{alfanu}) yields $\alpha_r (10d) = 0.91 \pm 0.14$, and 
       $\alpha_r (100d) = 1.17 \pm 0.11$. 
       The available measurements of the optical spectral slope, $\beta_{I-B}(3.8d) = 1.05 \pm 0.09$
       (Castro-Tirado \etal 2001), $\beta_o (7.5d)= 0.90 \pm 0.15$ (Djorgovski \etal 1999), and
       $\beta_{K-R}(8.5d) = 0.77 \pm 0.14$ (Bloom \etal 1998), indicate a hardening of the optical 
       emission that rules out the current scenario, in which it is expected that $\beta_o(8d) = 
       1.10 \pm 0.08$  
 \item 991216. For the $\beta_o$ given in Table 3 for Galactic extinction of $E(B-V)=0.63$, 
       the radio decay index is $\alpha_r (2d) = 0.13 \pm 0.12$, $\alpha_r (10d) = 0.52 \pm 0.09$,
       and $\alpha_r (100d) = 0.82 \pm 0.07$. However, the expected $X$-ray decay has an index
       $\alpha_x (0.1d) = 4.12 \pm 0.25$ and $\alpha_x (2d) = 2.47 \pm 0.10$, which is too large
       compared too observations. Similar results are obtained for $E(B-V)=0.46$ 
 \item 000301, for which $\alpha_r (30d) = 0.84 \pm 0.08$ and $\alpha_r (100d) = 1.26 \pm 0.08$.
       The optical slopes (corrected only for Galactic extinction) determined by Rhoads \& Fruchter (2001) 
       -- $\beta_{K-B}(2d) = 0.79 \pm 0.03$, $\beta_{K-B}(3d) = 1.04 \pm 0.03$, $\beta_{K-B}(5d) = 
       1.03 \pm 0.10$, $\beta_{K-R}(7.5d) = 0.69 \pm 0.10$, $\beta_{R-B}(13d) = 1.44 \pm 0.37$ -- 
       indicate a general softening with $\Delta \beta_o = 0.41 \pm 0.38$ from day 5 to day 13 
       (\ie after the optical light-curve break). The expected softening, $\Delta \beta_o = 0.21 \pm 
       0.04$, is consistent with the observations. 
\end{enumerate}
The evolution of $p$ required to yield the power-law decay of the optical emission of the afterglows
000926 and 010222 leads to radio fall-offs that are too steep in comparison with the observations. 

 Therefore, the scenario considered here is consistent with the available data in only one case 
(000301). We note that, a collimated outflows leads to a smaller $p_*$, as the effect of collimation 
is to yield steeper light-curve decays, and thus to larger radio decay indices (\eq [\ref{alfanu}]).
Consequently, a jet (spreading laterally or not), is even less able than a spherical outflow in 
explaining the shallow decay observed for the five anomalous afterglows.

\subsection{Time-Varying Electron and Magnetic Field Parameters}
\label{vareps}

 Below the absorption frequency $\nu_a$, the afterglow light-curve should should rise or be flat,
unless $\nu_a$ increased sufficiently fast. Such a behavior is not observed, thus the falling-off
radio emission rules out that spectral break located between radio and optical domains and having 
a non-standard evolution is $\nu_a$. If the break were the injection frequency $\nu_i$ then, 
to satisfy the condition $\nu_r < \nu_i$ until the last radio measurement requires large values 
for the $\epsB$ parameter. As shown in \S\ref{jet1} and \S\ref{fballwind}, in those cases where 
the shallow radio decay also requires that $\nu_c < \nu_i$, the resulting synchrotron peak flux 
is too large compared to the observed radio flux at about 100 days. As shown in Table 2, the
steepest radio fall-off that can be obtained for $\nu_r < \nu_i < \nu_c$, is $t^{-1/3}$ for a 
spreading jet. Thus, in the simplest model, the case $\nu_r < \nu_i$ cannot explain the steep 
radio decays observed for the above five anomalous afterglows, ranging from $t^{-2/3}$ to $t^{-1.2}$. 

 The remaining break to be considered is the cooling frequency $\nu_c$. Its location below the 
optical domain implies that the optical and $X$-ray light-curves must have the same decay, which is 
satisfied by the afterglow 000926, but is rather inconsistent with 991216 and 010222, for which 
$\alpha_x -\alpha_o = 0.21 \pm 0.08$ and $\alpha_x -\alpha_o = 0.12 \pm 0.04$, respectively. 
It is possible that the $X$-ray emission of these two afterglows has a significant contribution
from inverse Compton scatterings, which could yield a steeper fall-off that at optical frequencies.
For 991208 and 000301, there are no $X$-ray observations to test that $\nu_r < \nu_c < \nu_i$. 

 The non-standard evolution of $\nu_c$ can be determined from the difference $\Delta \alpha_{or}$ 
between the observed radio and optical decay indices. Given that the slope $\beta$ of the synchrotron 
spectrum, $F_\nu \propto \nu^{-\beta}$, steepens by $\delta \beta = 1/2$ from $\nu < \nu_c$ to 
$\nu > \nu_c$, it follows that $\alpha_o = \alpha_r - (1/2)(d\ln \nu_c/d\ln t)$, thus $\nu_c (t) 
\propto t^{-2(\alpha_o-\alpha_r)}$. The decay indices shown in Figure 1 lead to $\nu_c \propto 
t^{-1.6}$ for 991208, $\nu_c \propto t^{-1.8}$ for 991216, $\nu_c \propto t^{-2.8}$ for 000301, 
$\nu_c \propto t^{-3.2}$ for 000926, and $\nu_c \propto t^{-1.7}$ for 010222. Such steep fall-offs
of $\nu_c$ exceed by much that allowed in the simplest fireball model, where $\nu_c \propto t^{-1/2}$ 
for a homogeneous medium and $\nu_c \propto t^{1/2}$ for a wind-like medium (\eqs [\ref{nuc0}] and 
[\ref{nuc2}]). For a power-law density profile (\eq [\ref{nr}]) and a spherical outflow, the cooling 
frequency evolution is $d\ln \nu_c/d\ln t = (3s-4)/(8-2s)$. 
Thus, for $n$ increasing as a small power of the radius ($s < 0$), $\nu_c$ decreases roughly 
inversely proportional with $t$. The fastest fall-off of $\nu_c$ with time is reached for 
$s \rightarrow -\infty$, corresponding to a sharp increase in the external density, when $\nu_c 
\propto t^{-3/2}$, which is slightly slower than that required by the afterglows 991208, 991216, 
and 010222, and substantially slower than necessary for 000301 and 000926. 

 For this reason, we consider the case where there is a contribution to the evolution of $\nu_c$ 
arising from a time variation of the other afterglow parameters that determine it: $\epsB$ and
$\E$ (\eqs [\ref{nuc0}] and [\ref{nuc2}]). The magnetic field parameter $\epsB$ could vary in time 
if its value is determined by the Lorentz factor of the fireball, if the shocked external medium 
mixes with the GRB ejecta having a frozen-in magnetic field, or if the circumburst medium is 
magnetized. The fireball energy $\E$ may decrease in time due to radiative losses or may increase 
if the shock is refreshed through a delayed energy injection caused by a spread in the initial 
Lorentz factor of the ejecta (Rees \& \Meszaros 1998). Given that we want to explain both the 
observed radio and optical decay indices, $\alpha_r$ and $\alpha_o$, and that these indices depend 
on the electron parameter $\epsi$, we also allow this parameter to be variable. 
We note that radiative losses couple the variation of $\E$ to those of $\epsi$ and $\epsB$. 
However, to calculate this dependence, a knowledge of the entire electron distribution is 
necessary. For $p > 2$, it can be assumed that the distribution extends up to arbitrarily high 
energies, thus its slope and minimal electron energy (parameterized by $\epsi$) suffice, however
a high energy cut-off must exist for $p < 2$ (as is the case for 991216, 000301 and 010222, if 
$\nu_c < \nu_o$). For this reason, we allow the evolution of $\E$ to be independent of that of 
$\epsB$ and $\epsi$.

 The evolution of the afterglow parameters $\E$, $\epsi$, and $\epsB$ can be constrained with the
aid of the observed light-curve decay indices $\alpha_r$ and $\alpha_o$. To do this, we chose power-law 
dependences of the afterglow parameters on the observer time
\begin{equation}
  \E \propto t^e \;, \quad  \epsB \propto t^b \;, \quad  \epsi \propto t^i \;,
\label{plaws}
\end{equation}
for ease of calculations and because, in this way, a minimal number of unknowns is introduced. 
We note that, in a scenario where the afterglow parameters are time-varying, a steepening of 
the afterglow light-curve decay may be due to a non-uniform variation of the parameters, thus 
the optical light-curve breaks observed in the afterglows 991216, 000301, 000926 and 010222
may not require a tight collimation of the ejecta (\ie a jet). For this reason, we shall consider 
both spherical and collimated outflows.

\subsubsection{Spherical, Uniform Outflow}
\label{sph}

 In the assumption of a uniform temperature in the post-shock medium, equal to that set by the
shock-jump conditions, the fireball dynamics is given by $\Gamma^2 \M \propto \E$,
where $\M \propto r^{3-s}$ is the mass per solid angle of the energized external medium.
Together with $t \propto r/\Gamma^2$, it leads to 
\begin{equation}
 \log \Gamma = const + \frac{1}{8-2s} \log \E  - \frac{3-s}{8-2s} \log t \;, 
              \quad   r \propto (\E t)^{1/(4-s)} \;.
\label{dyn}
\end{equation}
Starting from the basic expressions of the spectral characteristics $F_p$, $\nu_i$, and $\nu_c$ 
(\eg Sari, Piran \& Narayan 1998), equation (\ref{dyn}) leads to
\begin{equation}
 \log F_p = const + \frac{8-3s}{8-2s} \log \E + \frac{1}{2} \log \epsB  - \frac{s}{8-2s} \log t \;,
\label{fp}
\end{equation}
\begin{equation}
 \log \nu_i = const + \frac{1}{2} \log\E + 2\log\epsi + \frac{1}{2}\log\epsB - \frac{3}{2}\log t \;, \quad 
 \log \nu_c = const - \frac{4-3s}{8-2s} \log \E - \frac{3}{2} \log \epsB - \frac{4-3s}{8-2s} \log t \;.
\label{nuic}
\end{equation}
For afterglow parameters varying as given in equation (\ref{plaws}), equations (\ref{spek}), 
(\ref{fp}), and (\ref{nuic}) yield the following light-curve decay indices at radio ($\nu_i < \nu_r 
< \nu_c$) and optical ($\nu_c < \nu_o$) frequencies:
\begin{equation}
 - \alpha_r = \left( \frac{p+3}{4} - \frac{s}{8-2s} \right) e + (p-1) i + \frac{p+1}{4} b 
            - \frac{3}{4}(p-1) - \frac{s}{8-2s}  \;,
\label{alphar}
\end{equation}
\begin{equation}
 - \alpha_o =  \frac{p+2}{4} e + \frac{p-2}{4} b + (p-1) i - \frac{3p-2}{4} \;.
\label{alphao}
\end{equation}
Our choice for $\nu_r > \nu_i$ is motivated by the uninterrupted power-law decay observed for the five 
anomalous radio afterglows until about 100 days, and the presence of a break in the radio light-curve 
of the afterglows 991208 (10 days), 000301 (30 days), and 000926 (20 days) that, most likely, is caused 
by the passage of $\nu_i$ through the radio domain. 

 Before using equations (\ref{alphar}) and (\ref{alphao}) and observations to constrain the evolution
of $\epsB$ and $\epsi$, let us determine the evolution of $\E$ in the absence of energy injection,
\ie the range of $e$ allowed by radiative losses. The fastest decrease of $\E$ thorough radiative losses
occurs when the shock-accelerated electrons acquire an energy equal to that in protons and cool
radiatively faster than adiabatically. In this fully radiative fireball, half of the energy is 
transferred to the circumburst medium during the time a mass equal to that already swept-up is
shocked, and is radiated away, \ie $\E \propto \M^{-1}$. Taking into account that the swept-up 
mass is $\M \propto r^{3-s}$, it yields $\E \propto r^{-(3-s)}$. Using $r$ from equation 
(\ref{dyn}), we obtain that, for a fully radiative fireball, $\E \propto \Gamma \propto t^{-(3-s)/(7-2s)}$ 
(see also \Meszaros, Rees \& Wijers 1998). Therefore, radiative losses are characterized by
\begin{equation}
  \E \propto t^e \quad {\rm with} \quad -\frac{3-s}{7-2s} \leq e \leq 0 \;.
\label{elimits}
\end{equation}

 For a given value of $i$, the exponents $e$ and $b$ can be determined from equations (\ref{alphar}) 
and (\ref{alphao}) and the electron index $p = 2\beta_o$ required by the slope of the afterglow optical 
continuum, $F_\nu \propto \nu^{-\beta_o}$. These slopes, corrected for Galactic extinction and possible 
Small Magellanic Cloud-like dust reddening in the host galaxy (inferred from the curvature of the optical 
spectrum under the assumption that the afterglow intrinsic continuum is a pure power-law), are given in 
Table 3. In the absence of energy injection (\ie for $e$ given in \eq [\ref{elimits}]), we reach the 
following conclusions for the five anomalous afterglows:
\begin{enumerate} 
 \item for $s=0$ and $e = -3/7$ (fully radiative fireball), $b > 0.9$ and $i < -1/4$.
       Taking into account that the magnetic field is $B \propto (n\epsB)^{1/2} \Gamma$ and
       that $\Gamma \propto t^{-3/7}$, it follows that $d\log B/d\log t > 0.02$
 \item for $s=0$ and $e = 0$ (adiabatic fireball), $b > 3/4$ and $i < -2/3$.
       From $\Gamma \propto t^{-3/8}$, it results that $d\log B/d\log t > 0$
 \item for $s=2$ and $e = -1/3$ (fully radiative fireball), $b > 4/3$ and $i < -1/3$.
       Then $\Gamma \propto t^{-1/3}$ and $r \propto t^{1/3}$ (\eq [\ref{dyn}]) lead to 
       $d\log B/d\log t > 0$
 \item for $s=2$ and $e = 0$ (adiabatic fireball), $b > 1.4$ and $i < -2/3$. 
       From $\Gamma \propto t^{-1/4}$ and $r \propto t^{1/2}$, it follows that
       $d\log B/d\log t > -0.05$.
\end{enumerate}  

 Therefore, for $s=0,2$ and without energy injection, $\epsi$ must decrease and $\epsB$ must
increase in time. The latter is sufficiently strong that leads to a magnetic field which is 
constant or increasing in time, which is too extreme for plausible mechanisms for 
generating a magnetic field (turbulence and mixing of magnetized ejecta with swept-up medium). 
For the afterglow 991208, the inferred behavior of $\epsi$ and $\epsB$ imply that, after 10 days,
the peak flux $F_p$ evolves as $t^0$ and $t^{0.3}$ for a radiative and an adiabatic outflow, 
respectively, while the injection frequency $\nu_i$ evolutions are $t^{-1.8}$ and $t^{-2.3}$. 
The resulting evolution of $F_p$ is rather inconsistent with the radio observations of 991208 at 
3--13 days -- $F_p \propto t^{-0.47 \pm 0.20}$ (Galama \etal 2001), while that of $\nu_i$ is in 
agreement with the observed $\nu_i \propto t^{-1.7 \pm 0.7}$.

 We note that, without energy injection, a slower increase of $\epsB$ is obtained for an increasing
external density. However, even in the limit of a sharp increase of the external density 
($s \rightarrow -\infty$), we obtain that $\epsB$ increases in time. In this case, for an
adiabatic fireball ($e=0$), the index $b$ for the five afterglows satisfies $b > 0.1$, implying
$d\log B/d\log t > -0.45$, while for a radiative outflow ($e=-1/2$), the resulting $b > 0.6$ 
leads to $d\log B/d\log t > -0.20$, \ie the magnetic field may decrease in time. For the afterglow 
991208, the resulting evolution of the peak flux ($F_p \propto t^0$ for a radiative outflow, 
$F_p \propto t^{0.5}$ for an adiabatic outflow) is not consistent with the observations, while 
that of the injection frequency ($\nu_i \propto t^{-1.8}$ and $\nu_i \propto t^{-2.7}$, respectively)
favors a radiative fireball.

 If there is energy injection ($e>0$), a constant parameter $\epsB$ can be obtained for $s=0$.
In addition, $\epsi$ must decrease for $s=0$, but it must increase for $s=2$. If the external
medium is homogeneous, the peak flux for the afterglow 991208 should increase faster than $t^{2.2}$ 
and the injection frequency should decrease faster than $t^{-5.8}$ at $t \simg 10$ days. This is 
in stark contrast with the radio observations of this afterglow (see above). For $s=2$, the 
parameter $\epsB$ increases in time even if there is an injection of energy.

\subsubsection{Non-Spreading Jet}
\label{coll}

 Before investigating a collimated outflow, we consider a structured outflow endowed with a core 
of a higher energy per solid angle $\E$ than outside it. In this case, the tangential motions are 
slow and the lateral spreading of the core is negligible (Kumar \& Granot 2003). Because the dynamics 
of the core remains unchanged, the spectral breaks continue to evolve as given in equation (\ref{nuic}). 
However, when the boundary of the core becomes visible to the observer, the increase of the area 
radiating the afterglow emission changes from $(r/\Gamma)^2$ to $(r \theta_c)^2$, where $\theta_c$
is the angular opening of the outflow core. This leads to a faster decrease of the peak flux 
corresponding to a multiplicative factor $\Gamma^2$. From equations (\ref{dyn}) and (\ref{fp}), 
it follows that when the core boundary is seen, the peak flux is
\begin{equation}
 \log F_p = const + \frac{10-3s}{8-2s} \log \E + \frac{1}{2} \log \epsB  - \frac{6-s}{8-2s} \log t \;,
\label{fpjet}
\end{equation}
and the index of the light-curve decay $F_\nu \propto t^{-\alpha}$ increases by 
\begin{equation}
 \delta \alpha = \frac{3-s-e}{4-s} 
\label{dalpha}
\end{equation}
at any frequency.
For an adiabatic fireball without energy injection ($e=0$), $\delta \alpha = 3/4$ for a homogeneous 
medium and $\delta \alpha = 1/2$ for a wind (Panaitescu, \Meszaros \& Rees 1998). Energy injection 
($e > 0$), decreases the steepening, while radiative losses ($e < 0)$ increase it. For the latter, 
the effect is small even for a fully radiative afterglow: $\delta \alpha = 6/7$ for $s=0$ and 
$\delta \alpha = 2/3$ for $s=2$. The optical afterglows 991216, 000301, 000926, and 010222 exhibited 
a break at around 1 day (Figure 1), the increase in the decay index being $\delta \alpha_o \sim 0.4, 
2.0, 0.7, 0.6$ (Table 3), respectively. Therefore the breaks seen in the optical light-curves of 
the afterglows 991216, 000926, and 010222 are consistent with those expected for a structured outflow 
endowed with a more energetic core, while that of the afterglow 000301 is too large and incompatible 
with a structured outflow. 
 
 As can be seen from equations (\ref{alphar}) and (\ref{alphao}), an additive term $\delta \alpha$
in the $rhs$ of each equation leads to a different solution for $i$ but leaves unchanged the value
of $b$. Therefore, in the scenario of a structured outflow without energy injection (\ie $e$ is 
constrained by \eq [\ref{elimits}]), the magnetic field strength must still be constant or increase 
in time to accommodate the radio and optical decay indices of the five anomalous afterglows, for the 
usual types of external media ($s=0,2$). However, in contrast to the uniform fireball case, a 
structured outflow without energy injection leads to evolutions of the peak flux and injection 
frequency that are consistent with the radio observations of the afterglow 991208 (Galama \etal 2001), 
for either a homogeneous or a wind-like circumburst medium. As for a uniform outflow, to obtain a
constant $\epsB$ for a non-spreading jet requires energy injection and leads to behaviors of $F_p$ 
and $\nu_i$ which are in conflict with the radio observations of 991208.

\subsubsection{Spreading Jet}
\label{jet}

 If the ejecta are confined to a jet with a sharp edge, its lateral spreading is not impeded and
the jet dynamics changes from that of a spherical outflow when this spreading becomes important.
Rhoads (1999) has shown that, for a homogeneous medium ($s=0$), the radius of a spreading jet 
increases only logarithmically with time, most of the increase of the swept-up mass being due to 
the lateral spreading. For an external density increasing with radius ($s < 0$), the jet radius
should increase even slower during the sideways expansion phase. However, the approximation of
a constant jet radius should become inadequate for a decreasing external density (Kumar \& 
Panaitescu 2001). For ease of calculations we make this approximation, which, for $s < 3$, has 
the effect of removing the dependence of the afterglow flux on the density profile of the external 
medium. 

 The dynamics of a spreading jet is given by (Rhoads 1999)
\begin{equation}
 \Gamma \propto E^{1/6} t^{-1/2} \;, \quad M \propto E^{2/3} t \;, 
\label{dynjet}
\end{equation}
where $E$ is the jet energy and $M$ is the mass of swept-up external matter.
From here, we obtain the following expressions for the spectral characteristics:
\begin{equation}
 F_p \propto E^{4/3} \epsB^{1/2} t^{-1} \;, \quad \nu_i \propto E^{2/3} \epsi^2 \epsB^{1/2} t^{-2} \;,
     \quad \nu_c \propto E^{-2/3} \epsB ^{-3/2} t^0 \;.
\end{equation}
Substituting the parameter evolution given in equation (\ref{plaws} in the resulting afterglow 
light-curves, yields the following decay indices
\begin{equation}
  - \alpha_r = \frac{p+3}{3} e + \frac{p+1}{4} b + (p-1) i - p  \;, 
     \quad {\rm and} \quad
  - \alpha_o = \frac{p+2}{3} e + \frac{p-2}{4} b + (p-1) i - p  \;.
\end{equation}

 The dynamics of a fully radiative jet can be derived from equation (\ref{dynjet}), taking into
account that $E \propto M^{-1}$ (half of energy is radiated away while the swept-up mass doubles).
Then $E \propto \Gamma \propto t^{-3/5}$. Therefore, in the absence of energy injection, radiative 
losses yield $-3/5 \leq e \leq 0$. From equations (\ref{alphar}) and (\ref{alphao}), we obtain that 
the observed decay indices of the five anomalous afterglows require
\begin{enumerate}
 \item $b > 4/3$ for $e = -3/5$ (fully radiative jet). From $B \propto \epsB^{1/2} \Gamma$
       and $\Gamma \propto t^{-3/5}$, it follows that $d\log B/d\log t > 0.07$
 \item $b > 1.1$ for $e = 0$ (adiabatic jet). Using $\Gamma \propto t^{-1/2}$, it yields 
       $d\log B/d\log t > 0.05$
\end{enumerate}
Therefore, for $s=0,2$ and without energy injection, $\epsB$ must increase in time and, just as for 
a spherical outflow, this increase is sufficiently strong that it leads to a magnetic field that is 
constant or increases with time. In contrast, the electron parameter $\epsi$ may either decrease or
increase. The resulting behavior of the peak flux $F_p$ and injection frequency $\nu_i$ for the
afterglow 991208 are in agreement with the observations if the jet is adiabatic. To obtain a constant 
$\epsB$, $\epsi$ must be decrease faster than $t^{-2.5}$, which, for the afterglow 991208, leads to 
the same inconsistencies between the evolutions of $F_p$ and $\nu_i$ and observations that were 
mentioned for a spherical outflow (\S\ref{sph}).

\subsection{Structured Outflow}
\label{structure}

 So far, we have considered two scenarios where the radio and optical afterglows arise from the
same population of electrons. If there is no break in the synchrotron spectrum between the radio 
and optical domains and if the slope of the electron energy distribution does not change in time, 
then the discrepancy between the radio and optical decays of the afterglows 991208, 991216, 000301, 
and 000926 must be explained by that these emissions arise from different regions of the GRB outflow. 
The key quantity in the calculation of the afterglow emission is the outflow Lorentz factor, which 
is determined by the energy per solid angle and the degree of collimation of each region. Thus, one 
can imagine that, in a structured outflow, a part of it has a higher kinetic energy and a larger 
Lorentz factor than another part, so that the former radiates more in the optical and less in the 
radio than the latter. 

 Within the structured outflow scenario, an optical decay steeper than in the radio can be obtained 
if the outflow radiating in the optical is tightly collimated. Then, for an adiabatic fireball and 
without energy injection, the optical decay index increases by $\delta \alpha = 3/4$ for $s=0$ and 
by $\delta \alpha = 1/2$ for $s=2$ at the time $t_b$ when the boundary of the optically emitting 
outflow becomes visible (\eq [\ref{dalpha}]). As discussed in \S\ref{coll}, the $\delta \alpha$ 
expected for a structured outflow is consistent with that observed for the afterglows 991216, 000926, 
and 010222, but not for 000301. Furthermore, the near-infrared--optical spectral slope determined 
for the afterglow 000301 by Jensen etal (2001), $\beta_o = 0.57 \pm 0.02$, even the steepest
post-break decay index that can be obtained with a structured outflow, $\alpha_o = 1.5 \beta_o +1
= 1.86 \pm 0.03$, is too shallow compared with that observed. For these reasons, this afterglow
will not be considered in this section.

 If the radio emitting outflow is quasi-spherical, then the differential decay $\Delta \alpha_{or} = 
\alpha_o(t>t_b) - \alpha_r$ is equal to $\delta \alpha$ if the cooling frequency of the jet emission 
is blueward of the optical and to $\delta \alpha + 1/4 = 1$ ($s=0$) or $\delta \alpha - 1/4 = 1/4$ ($s=2$) 
if it is below. For the afterglows 991208, 991216, and 010222, $\Delta \alpha_{or} = 0.8 \pm 0.1$, 
$0.9 \pm 0.1$, and $0.84 \pm 0.05$, respectively, thus the differential decay seen between the radio 
and optical emission of the first three afterglows may be accommodated by an optical jet interacting 
with a homogeneous medium, particularly if the cooling frequency is redward of the optical domain.
However, the $\Delta \alpha_{or} = 1.6 \pm 0.1$ of the afterglow 000926 exceeds that allowed in 
this scenario, thus allowance for a different electron index $p$ of the radio and optical outflows 
must be made. For generality, we will allow for different indices $p$ in the radio and optical 
outflows of all afterglows, thus the only remaining restriction is that of a homogeneous medium for 
the afterglow 000926, where the observed $\delta \alpha = 0.7$ is inconsistent with a wind-like medium.

 Having outlined the general set-up for a scenario that accounts for the observed radio and optical
decays (for four of the five anomalous afterglows) -- a structured outflow containing a core with
a higher energy per solid angle, producing the optical afterglow, and yielding a light-curve break 
when its edge becomes visible, and an envelope of lower energy per solid angle, from where the 
radio afterglow arises -- we proceed to constrain this scenario by requiring that the optical 
and radio outflows dominate the emission in their respective domain over the entire time interval 
when the afterglow decay is observed. These constraints are suggested by that, as it softens, the 
synchrotron emission from the optical core yields a radio contribution that falls-off shallower 
than that from the radio envelope, while the slower deceleration of the radio envelope yields an 
optical emission that decays slower than that from the optical core. 

  For simplicity, we consider that the angular distribution of the ejecta energy is uniform in 
both the core and the envelope of the relativistic outflow, \ie the fireball angular structure 
is a top-hat function (co-axial uniform outflows). Such a "dual fireball" scenario (Frail \etal
2000) yields the maximal decoupling of the radio and optical emissions.

 That the optical light-curve decays from the earliest observation indicates that the line 
observer--fireball center is within the optical jet. In this case, the radio envelope, being 
less relativistic than the optical core, becomes visible to the observer before the edge of the 
core, thus the radio light-curve break seen in the afterglows 991208 and 000926 at $t_r = 10-20$ 
days, well after the time of the optical break, cannot be due to the outflow structure and 
relativistic effects. Given that for plausible afterglow microphysical parameters the injection 
frequency $\nu_i$ is expected to cross the radio domain at around 10 days (\eq [\ref{nui}]), it
is more natural to associate the radio light-curve breaks of 991208 and 000926 (and of all other
radio afterglows in Figure 1) with the passage of $\nu_i$ through the radio band.

 Assuming that the parameters $\epsi$ and $\epsB$ are the same everywhere, the larger energy per
solid angle within the core implies that its injection frequency $\nuiO$ is higher than for the 
envelope, therefore $\nuiO$ crosses the radio domain at a time $t_+ > t_r$. For $t_b < t < t_+$, 
the radio emission from the optical jet is that for a spherical outflow, $F_\nu \propto t^{1/2}$ 
for $s=0$ or $F_\nu \propto t^0$ for $s=2$, corrected for the frequency-independent steepening 
$\delta \alpha$ (=3/4 for $s=0$, =1/2 for $s=2$) produced by seeing the jet edge, therefore 
$\FrO \propto t^{-1/4}$ for $s=0$ and $\FrO \propto t^{-1/2}$ for $s=2$. These fall-offs are 
shallower than those observed in all radio afterglows (with the exception of 010222 for $s=2$), 
which we attribute to the radio envelope. For $t > t_+$, the decay index of the radio emission 
from the optical jet is equal to that of the optical light-curve or differs by 1/4, thus 
$\FrO (t > t_+)$ decays steeper than the observed radio light-curves. Therefore, the dominance 
of the radio emission from the radio envelope $\FrR(t)$ is ensured at all times if it is brighter 
than the radio flux from the optical jet at $t=t_+$: $\FrR(t_+) > \FrO (t_+)$. From $t_+ > t_r$, 
it follows that  
\begin{equation}
 \FrR (t_+) = \Frtr \left( \frac{t_+}{t_r} \right)^{-\alpha_r} \;,
\label{FrR}
\end{equation}
where $\Frtr$ is the observed radio flux at the time $t_r \sim 10$ days when the radio fall-off 
starts, though any subsequent time can be chosen if the onset of the radio decay has been missed.
For $t_b < t < t_+$, the radio emission from the optical jet falls-off as $\FrO \propto t^{-1/(4-s)}$
(see above), therefore
\begin{equation}
 \FrO (t_+) = \FpOtb \left[ \frac{\nu_r}{\nuiOtb} \right]^{1/3} \left( \frac{t_+}{t_b} \right)^{-1/(4-s)} \;, 
\label{FrO}
\end{equation}
where $\FpOtb$ is the peak synchrotron flux from the optical jet at $t=t_b$. This quantity 
can be determined from the observed optical flux $\Fotb$ which, taking into account that $\nuiOtb 
< \nu_o$, is
\begin{equation}
  \Fotb = \FpOtb \left[ \frac{\nuiOtb}{\nu_o} \right]^{(p_O-1)/2} 
          \min \left\{ 1, \left[ \frac{\nucOtb}{\nu_o} \right]^{1/2} \right\} \;,
\label{Fotb}
\end{equation}
where $p_O$ is the electron index in the optical jet. From that $\nuiO (t) \propto t^{-3/2}$
at all times and for any $s$, the crossing time $t_+$ defined by $\nuiO (t_+) = \nu_r$ is 
$t_+ = t_b [\nuiOtb/\nu_r]^{2/3}$. Substituting it in equations ({\ref{FrR}}) and (\ref{FrO}) and
using equation (\ref{Fotb}), the above condition $\FrR(t_+) > \FrO(t_+)$ yields a lower limit on
$\nuiOtb$:
\begin{equation}
 \left[ \frac{\nuiOtb}{\nu_r} \right]^\xi > \frac{\Fotb}{\Frtr} \left(\frac{t_b}{t_r}\right)^{\alpha_r} 
                                            \left(\frac{\nu_o}{\nu_r}\right)^{(p_O-1)/2} 
                              \max \left\{ 1, \left[ \frac{\nu_o}{\nucOtb} \right]^{1/2}  \right\} \;,
      \quad {\rm where} \quad
 \xi \equiv \frac{1}{2}(p_O-1) + \frac{2-s/3}{4-s} - \frac{2}{3}\alpha_r \;.
\label{nuiO}
\end{equation}

 From equation (\ref{nuiO}) it can be shown that the least stringent constraint on $\nuiOtb$ is 
obtained if $\nucOtb > \nu_o$. Physically, this can be understood from that, for a given optical flux, 
the peak flux of the core emission must be larger for $\nucOtb < \nu_o$ than for $\nucOtb > \nu_o$. 
A larger peak flux leads to brighter radio emission from the optical jet at $t=t_+$.
To compensate, a larger $t_+$ is required because $\FrO(t)$ decreases in time. Finally, a larger $t_+$
implies a higher $\nuiOtb$. As we shall see, the higher is the lower limit on $\nuiOtb$, the more it
becomes a problem for the scenario considered here, thus we restrict now our attention 
to the less constraining case $\nucOtb > \nu_o$, for which $\beta_o = (p_O-1)/2$. 
Because the observed slope of the optical afterglow is affected by an uncertain extinction in the host 
galaxy, we choose to determine the index $p_O$ from the optical decay index $\alpha_o (t > t_b)$, which 
is better constrained by observations than that before the break. When the jet edge is seen, the optical
light-curve at $\nu_o < \nu_c$ steepens from $F_o \propto t^{(3p-3)/4}$ to $F_o \propto t^{3p/4}$ for $s=0$, 
and from $F_o \propto t^{(3p-1)/4}$ to $F_o \propto t^{(3p+1)/4}$ for $s=2$. Thus $p_O = (4/3) \alpha_o$ 
for $s=0$ and $p_O = (4\alpha_o-1)/3$ for $s=2$. Substituting $p_O$ in the above expression of $\xi$
leads to $\xi = (2/3) (\alpha_O - \alpha_R)$ for either type of external medium.

\vspace*{2mm}
 Table 3 lists the values of the observables that appear in the $rhs$ of equation (\ref{nuiO}). 
Substituting them, we obtain for $\nu_o < \nucOtb$ the following results:
\vspace*{-2mm}
\begin{enumerate}
\item GRB 991208. 
       For $s=0$, $p_O = 2.37 \pm 0.12$, slope of intrinsic optical spectrum $\beta_O = (p_O-1)/2 = 
       0.69 \pm 0.06$ (requiring $A_V^{host} \sim 0.2$ to be consistent with the slope determined by 
       Castro-Tirado \etal 2001), 
       $\nuiOtb > 8,000$ GHz. 
       For $s=2$, $p_O = 2.04 \pm 0.12$, $\beta_O = 0.52 \pm 0.06$ (requiring $A_V^{host} \sim 0.3$), 
       $\nuiOtb > 240$ GHz.
\item GRB 991216. 
       For $s=0$, $p_O = 2.12 \pm 0.07$, $\beta_O = 0.56 \pm 0.03$ (consistent with the slope measured 
          by Garnavich \etal 2000 for a Galactic extinction of $E(B-V)=0.63$), 
       $\nuiOtb > 6,000$ GHz.
       For $s=2$, $p_O = 1.79 \pm 0.07$, $\beta_O = 0.40 \pm 0.03$ (requiring $A_V^{host} \simg 0.1$), 
       $\nuiOtb > 340$ GHz. 
\item GRB 000926. 
       For $s=0$, $p_O = 3.07 \pm 0.03$, $\beta_O = 1.03 \pm 0.01$ (consistent with the value inferred 
          by Fynbo \etal 2001 after correction for host extinction of $A_V^{host} \sim 0.2$), 
       $\nuiOtb > 5,000$ GHz. 
\item GRB 010222.
       For $s=0$, $p_O = 1.69 \pm 0.01$, $\beta_O = 0.35 \pm 0.01$ (requiring $A_V^{host} \sim 0.2$ 
          to be consistent with the slope determined by Jha \etal (2001) and Stanek \etal (2001), 
          thus exceeding the host extinction inferred by Galama \etal 2003 and Lee \etal 2003, 
          $A_V^{host} \siml 0.1$), $\nuiOtb > 730$ GHz. 
       For $s=2$, $p_O = 1.36 \pm 0.01$, $\beta_O = 0.18 \pm 0.01$ (requiring $A_V^{host} \sim 0.3$), 
       $\nuiOtb > 30$ GHz. 
\end{enumerate}
If the above values of $\nuiOtb$ are extrapolated back to the time $t_{min}$ of the first optical 
measurement, the resulting $\nuiO(t_{min})$ is redward of the optical domain, consistent with that
the optical light-curves decay from the first observation.
We note that, for a homogeneous medium, the electron index required by the observed radio decay, 
$p_R = (4\alpha_r + 3)/3$, are compatible with the above values of $p_O$ for all afterglows except
000926. However, for a wind-like medium, the index $p_R = (4\alpha_r + 1)/3$ is always smaller
than $p_O$. Thus, the latter type of circumburst medium requires different electron indices in the
outflow envelope and core, which makes this scenario somewhat contrived.
        
 The above lower limit on $\nuiOtb$ can be converted into a lower limit on the injection frequency
$\nuiRtr$ for the radio component at the onset of the radio decay. As argued before, we expect 
that $\nuiRtr = 10$ GHz (or $\nuiRtr < 10$ GHz, if the beginning of the radio decay occurred before
the first observation), thus a self-consistency test for the structured outflow scenario is to require 
that the lower limit on $\nuiRtr$ is below the radio domain. From equation (\ref{nui}), we obtain 
\begin{equation}
  \log\frac{\nuiRtr}{\nuiOtb} = \frac{1}{2} \log\frac{\ER}{\EO} - \frac{3}{2} \log\frac{t_r}{t_b} \;,
\label{nuratio}
\end{equation} 
where $\ER$ and $\EO$ are the energy densities of the radio and optical outflows, respectively.
This ratio can be determined from the observed radio and optical fluxes ($\Frtr$ and $\Fotb$) and
equations (\ref{nui}), (\ref{fp0}), and (\ref{fp2}). At an observing frequency above $\nu_i$ and
below $\nu_c$, the afterglow emission is $F_\nu (t)= F_p (\nu/\nu_i)^{-\beta}$ (\eq [\ref{spek}]), 
therefore
\begin{equation} 
 \frac{\Frtr}{\Fotb} = \frac{F_p^{(R)}(t_r)}{\FpOtb} \left[\frac{\nuiRtr}{\nu_r}\right]^{\beta_r}
                                             \left[\frac{\nu_o}{\nuiOtb}\right]^{\beta_o} \;,
\label{Fratio}
\end{equation}
where $\beta_r = (p_R-1)/2$ is the slope of the radio spectrum, $p_R$ being the electron index of the
radio outflow, which can be determined from the observed decay index: $\alpha_r = (3p_R-3)/4$ for $s=0$ 
and $\alpha_r = (3p_R-1)/4$. By substituting $\nuiRtr$ from equation (\ref{nuratio}) and $F_p$ from 
equation (\ref{fp}), we obtain
\begin{equation}
  \log\frac{\Frtr}{\Fotb} = \left(\frac{8-3s}{8-2s}+\frac{\beta_r}{2} \right) \log\frac{\ER}{\EO} +
                            \left(\frac{s}{8-2s}+\frac{3\beta_r}{2}\right) \log\frac{t_b}{t_r} +
                            \beta_r \log\frac{\nuiOtb}{\nu_r} + \beta_o \log\frac{\nu_o}{\nuiOtb} \;.
\end{equation}
Extracting the ratio $\ER/\EO$ from here and substituting it in equation (\ref{nuratio}), yields
\begin{equation}
   \left(\frac{8-3s}{4-s}+\beta_r\right) \log\frac{\nuiRtr}{\nuiOtb} =  
                  \log\frac{\Frtr}{\Fotb} - \left(\frac{12-5s}{4-s}\right) \log\frac{t_b}{t_r} + 
                  \beta_r \log\frac{\nu_r}{\nuiOtb} + \beta_o \log\frac{\nuiOtb}{\nu_o} \;. 
\label{nuiR}
\end{equation}

 Note from the above equation that $\nuiRtr \propto [\nuiOtb]^{(2+\beta_o)/(2+\beta_r)}$ for $s=0$
and $\nuiRtr \propto [\nuiOtb]^{(1+\beta_o)/(1+\beta_r)}$ for $s=2$. Because $\nuiRtr$ increases with
increasing $\nuiOtb$, it follows that the lower limit on $\nuiOtb$ given in equation (\ref{nuiO})
sets a lower limit on $\nuiRtr$. For the afterglow characteristics given in Table 3, and with $\beta_r$
and $\beta_o$ determined from the decay indices $\alpha_r$ and $\alpha_o$, respectively, equation 
(\ref{nuiR}) leads to the following results
\begin{enumerate}
 \item GRB 991208: $\nuiRtr > 200$ GHz for $s=0$ and  $\nuiRtr > 9$ GHz for $s=2$,
 \item GRB 991216: $\nuiRtr > 530$ GHz for $s=0$ and  $\nuiRtr > 40$ GHz for $s=2$,
 \item GRB 000926: $\nuiRtr >  26$ GHz for $s=0$,
 \item GRB 010222: $\nuiRtr > 130$ GHz for $s=0$ and  $\nuiRtr > 9$ GHz for $s=2$.
\end{enumerate}
Therefore, we find that the condition that the radio emission from the envelope outflow is brighter 
than that from the optical jet leads to that $\nuiRtr$ is above the radio domain instead of the
$\nuiRtr \simeq \nu_r$ expected from the decay of the radio afterglows after $t_r$. Only for two GRB 
afterglows (991208 and 010222), the structured outflow scenario may be compatible with the observations, 
provided that $s=2$, which, as discussed above, requires different electron indices in the outflow
core and envelope.  

 In the $\nucOtb < \nu_o$ case, an even higher lower limit on $\nuiRtr$ is suggested by that 
$\nuiOtb$ is now larger. However, for a self-consistent treatment, the result given in equation
(\ref{nuiR}) should be re-derived after adjusting equation (\ref{Fratio}) for the $\nucOtb > \nu_o$
case. The additional constraint on $\nucOtb$ necessary to calculate the upper limit on $\nuiRtr$
is provided by requiring that the optical emission from the radio envelope does not overshine that 
from the optical jet at any time when the power-law fall-off of the optical afterglow is observed, 
otherwise a flattening of the optical light-curve is expected\footnote{
 Castro-Tirado \etal (2001) note such a behavior in the afterglow 991208, at 20--30 days. 
 Apart from a substantial contribution from the radio outflow, fluctuations in the circumburst 
 medium density, an episode of energy injection, or the emergence of an associated supernova, 
 are other plausible explanations for the late-time bump in the optical afterglow 991208} .
This constraint is satisfied at all times if it is fulfilled at the time $t_{max}$ of the latest 
optical measurement. Similar to the above calculation of the radio emission from the optical core
and the radio envelope, the condition $F_o^{(O)}(t_{max}) > F_o^{(R)}(t_{max})$ leads to an upper 
limit on the cooling frequency $\nucRtr$ for the radio envelope emission.
From ${\ER} < {\EO}$ and $\nu_c \propto \E^{1/2}$ for a homogeneous medium, it follows that, for 
this type of external medium, $\nucRtr$ is also an upper limit on $\nucO(t_r)$, the 
cooling frequency for the optical core emission, which can be easily converted into an upper limit
on $\nucOtb$.

 In the $\nucOtb < \nu_o$ case, the analytical calculation of $\nuiRtr$ is cumbersome and it is
best to convert all constraints on the afterglow spectral properties into constraints on the afterglow
model parameters. The five parameters $\ER$, $\EO$, $n$ (or $A_*$ for $s=2$), $\epsi$, and $\epsB$ for 
a structured outflow are constrained by three equations (for $\nuiRtr = \nu_r$, $\Frtr$, and $\Fotb$) 
and two inequalities (for $\nucRtr$ and $\nuiOtb$). Solving for the allowed parameter ranges
in all cases above where $\nuiRtr \sim 10$ GHz, we find that $\ER < 10^{50}$ ergs, $\epsi > 1/4$, and 
$A_* > 20$ (for 991208), $A_* > 150$ (for 991216), $n \simg 10^5 \cm3$ (for 000926), and $A_* \simg 200$ 
(for 010222). Such large external densities, characteristic for a red supergiant or a molecular cloud, 
lead to a semi-relativistic motion ($\Gamma < 1.5$) and a self-absorbed radio emission at $t > t_r$. 
More worrisome, the electron parameter $\epsi$, together with the electron distribution slope $p_R < 2$ 
required by the observed radio decay index $\alpha_r$, imply total electron energies that exceed 
equipartition with protons. Thus, the marginally viable cases require implausible afterglow parameters.

\subsection{Reverse Shock in Ejecta Outflow}
\label{reverse}

 In this section we consider the possibility that the slowly decaying radio emission observed in some 
GRB afterglows arises in a long-lived reverse shock crossing some incoming, delayed ejecta, while the 
optical emission is produced in the forward shock which energizes the circumburst medium. An injection 
of ejecta in the fireball, lasting for the entire duration of the radio observations (tens to hundreds 
of days), is unlikely to be due to a very long-lived central engine. Instead we consider that the
delayed injection is caused by that the ejecta are expelled by the GRB progenitor with a broad range of 
initial Lorentz factors (Rees \& \Meszaros 1998), the slower ejecta lagging behind the faster part of 
the outflow and catching-up with it during the afterglow phase, as the leading edge of the fireball is 
decelerated by the circumburst medium.

 Assuming constant parameters $\epsi$ and $\epsB$, the steepening of the optical light-curve decay must 
be associated with a tight collimation of the ejecta because a sudden reduction of the injected power 
cannot explain the steepness of the decay after the break time $t_b$. For a jet with sharp edges,
the slower moving, leading edge of the outflow spreads laterally at $t > t_b$ faster than the
incoming ejecta, which leads to angular non-uniformities in the Lorentz factor of the fireball.
To avoid this complication, we assume that the jet does not have sharp edges, \ie it is embedded
in an envelope (as considered in \S\ref{structure}) which prevents its lateral spreading, but
yielding a negligible contribution to the afterglow emission.


 The decay of the radio emission from the reverse shock is determined by the evolution of $\nuiRS$ 
and of the peak flux $F_p^{(R)}$, which depend on the distribution of energy with Lorentz factor
in the ejecta and on the Lorentz factor $\Gamma_i(t)$ of the ejecta which catch-up with the leading
edge of the fireball at some observer time $t$. For the dynamics of the latter, we consider the simple 
law $\E \propto t^e$ for the fireball energy. Most likely, the ejecta are expelled over a timescale 
much shorter than the observer time ($\simg 0.1$ day) when the afterglow observations are made. 
Because of this, the Lorentz factor $\Gamma_i$ of the freely expanding ejecta which enter the afterglow 
fireball at $t \simg 0.1$ days is independent on the duration of the ejecta release. From the kinematics 
of the ejecta--fireball catch-up, one obtains that
\begin{equation}
 \frac{\Gamma_i}{\Gamma} = \left( \frac{4-s}{e+1} \right)^{1/2} \quad (e < 3-s) \;.
\label{Gmi}
\end{equation} 
where $\Gamma$ is the Lorentz factor of the decelerating fireball (\eq [\ref{dyn}]). Since the 
fireball Lorentz factor at 100 days is less than a few, the above equation shows that, in order
for the reverse shock to last for $\sim 100$ days, the GRB progenitor must release ejecta whose
initial Lorentz factor is $\Gamma_i < 3\div 5$. 
We note that the above ratio of Lorentz factors is time-independent. Because the Lorentz factor of 
the reverse shock in the comoving frame of the unshocked ejecta depends on this ratio, it follows 
that the reverse shock Lorentz factor and the typical electron energy behind the reverse shock do 
not evolve in time. 
Then, if the magnetic parameter $\epsB$ is also the same (which implies that the magnetic field is 
identical), the reverse shock injection frequency evolves as $\nuiRS \propto \nuiFS/\Gamma^2$, 
where $\nuiFS$ is the injection frequency behind the forward shock.

 The equality of $\epsB$ behind both shocks also implies that the ratio of the reverse to 
forward shock peak fluxes is just the ratio of the corresponding number of emitting electrons,
$F_p^{(R)}/F_p^{(F)} = \Mi/\M$, where $\Mi$ and $M$ are the ejecta mass and that of the swept-up 
circumburst medium, respectively. For simplicity, we assume that the mass of the ejecta energized by 
the reverse shock increases as a power-law in the Lorentz factor of the incoming ejecta: 
\begin{equation}
 \Mi (>\Gamma_i) \propto \Gamma_i^{-q} \;,\; q > 0 \;,
\label{Mi}
\end{equation}  
therefore $\Mi \propto \Gamma^{-q}$ (see \eq [\ref{Gmi}]). Equation (\ref{Mi}) is equivalent to the
assumption that the energy distribution in the ejecta is $d\E_i/d\Gamma_i \propto \Gamma_i^{-q}$.  

 Using equations (\ref{nuic}) and (\ref{fpjet}) with $\E \propto t^e$, the evolution of the reverse 
shock injection frequency and peak flux are
\begin{equation}
 \frac{d\log \nuiRS}{d\log t} = - \frac{6-s - e(2-s)}{8-2s} \;, \quad
 \frac{d\log F_p^{(R)}}{d\log t} = -\frac{1}{2}(3-e) + \frac{3-e-s}{8-2s} q \;. 
\label{RS}
\end{equation} 
From here it can be shown that the radio flux from the reverse shock, $F_r^{(R)} \propto F_p^{(R)}
(\nuiRS)^{(p-1)/2}$, evolves as $F_r^{(R)} \propto t^{-\alpha_r}$ with 
\begin{equation}
 \alpha_r = \frac{3-e}{2} + \frac{6-s-e(2-s)}{4(4-s)}(p-1)- \frac{3-e-s}{2(4-s)} q \;,
\label{ar}
\end{equation}
where $p$ is the electron index for the reverse shock, which we assume to be the same as for the forward
shock.
 The exponent $e$ which quantifies the amount of injected energy can be estimated from the observed
optical light-curve decay index $\alpha_o (t > t_b)$. From equations (\ref{nuic}), (\ref{spek}),
and (\ref{fpjet}), it can be shown that the decay index of the forward shock optical light-curve is
\begin{equation}
  \alpha_o = \frac{3-e}{4} p - \frac{e(16-5s)-s}{4(4-s)} \;\; {\rm if} \;\;\nu_o < \nu_c  \quad {\rm and} 
    \quad 
  \alpha_o = \frac{3-e}{4} p - \frac{e(6-s)+s-2}{2(4-s)} \;\; {\rm if} \;\;\nu_c < \nu_o \;, 
\label{ao}
\end{equation}
where $p$ is the electron distribution index for the forward shock, which can be determined from the
observed optical spectral slope $\beta_o$.

 Equations (\ref{ar}) and (\ref{ao}) and the observed decay indices (Table 3) allow the determination
of the indices $e$ and $q$. Generally, there are three constraints that these indices must satisfy. 
First, if the injected energy ($\E_i (> \Gamma_i)$) exceeds that existing initially ($\E_0$), then the 
exponents $q$ and $e$ are not independent. From $\E \simeq \E_i \propto \Gamma_i^{1-q}$, $\E \propto t^e$ 
and equation (\ref{dyn}), one obtains 
\begin{equation}
 e = \frac{(3-s)q}{7-2s+q} \;.
\label{eq}
\end{equation}
However, an energy injection that is dynamically unimportant ($\E_i (> \Gamma_i) \ll \E_0$) does not 
necessarily imply a negligible injected mass: in the $e \rightarrow 0$ limit, $e$ and $q$ are independent 
quantities. Given that the radio shallow decays are observed to last for 1--2 decades in time, the
constraint given in equation (\ref{eq}) should be used only for $e> 1/3$, corresponding to at least 
a two-fold increase of the fireball energy.
The second constraint is that the light-curve steepening $\delta \alpha (e)$ given in equation 
(\ref{dalpha}), resulting from seeing the core edge, must be larger than that observed in the optical, 
$\delta \alpha_o$ (Table 3). The reason is that, if measurements did not capture the asymptotic 
early-time optical light-curve index, then the $\delta \alpha_o$ determined through power-law fits to 
the pre- and post-break emission is an underestimation of the true index increase $\delta \alpha(e)$.
The third constraint is that $e$ must exceed the lowest value allowed by radiative losses: 
$e \geq -(3-s)/(7-3s)$.

 Imposing the above constraints for the indices $e$ and $q$, we find that the energy injection scenario 
is compatible with the observations in the following cases:
\begin{enumerate}
 \item GRB 991208, 
   \begin{itemize} \vspace*{-2mm}
    \item $s=0$ and $\nu_o < \nu_c$ (with $e = 0.3$, $q = 3.2$),
    \item $s=0$ and $\nu_c < \nu_o$ ($e = 0$, $q = 2.5$),
    \item $s=2$ and $\nu_c < \nu_o$ ($e = -0.1$, $q = 4.0$),
   \end{itemize}
  \vspace*{-1mm}
    leading to reverse shock peak fluxes evolutions (\eq [\ref{RS}]) which are consistent with those measured 
    by Galama \etal (2003), but to injection frequencies whose evolutions are marginally consistent
    with the observations, 
  \vspace*{1mm}
 \item GRB 991216, for $\beta_o = 0.58$
   \begin{itemize} \vspace*{-2mm}
    \item $s=0$ and $\nu_o < \nu_c$ ($e = 0$, $q = 3.3$),
    \item $s=0$ and $\nu_c < \nu_o$ ($e = 0$, $q = 6.6$),
   \end{itemize}
   \vspace*{-1mm} \hspace*{1cm} while for $\beta_o = 0.87$  
   \begin{itemize} \vspace*{-2mm}
    \item $s=0$ and $\nu_o < \nu_c$ ($e = 0.3$, $q = 3.7$),
    \item $s=0$ and $\nu_c < \nu_o$ ($e = 0$, $q = 2.9$),
    \item $s=2$ and $\nu_c < \nu_o$ ($e = -0.2$, $q = 4.3$),
   \end{itemize}
 \item GRB 000926, $s=0$ and $\nu_o < \nu_c$ ($e = 0$, $q=4.2$),
  \vspace*{1mm}
 \item GRB 010222, 
   \begin{itemize} \vspace*{-2mm}
    \item $s=0$ and $\nu_c < \nu_o$ ($e = 0.1$, $q=3.4$),
    \item $s=2$ and $\nu_c < \nu_o$ ($e = -0.1$, $q=5.2$).
   \end{itemize}
\end{enumerate}
For the slope of the optical continuum determined by Rhoads \& Fruchter (2001), the sharp fall-off 
of the optical emission of 000301 is exceeds that allowed by the core edge becoming visible.
Alternatively, that steep decay would require the outflow energy to decrease faster than allowed
by radiative losses. Therefore, the properties of the afterglow 000301 cannot be accommodated
by the current scenario, and will be left out for the remainder of this section.

 The above values of the $e$ parameter show that the energy of the ejecta reaching the forward 
shock over a tenfold increase in observer time is at most comparable to the initial ejecta energy. 
This is due to that, for the observed optical spectral slope $\beta_o$, a significant energy injection 
reduces the optical light-curve steepening expected when the core edge becomes visible (\eq [\ref{dalpha}]) 
below that which is observed. For generality, we maintain the parameter $e$ in the following equations,
although $e \simeq 0$ for the four anomalous afterglows.

 As a further test for the reverse--forward shock scenario, we require that the forward shock emission
is dimmer than that from the reverse shock emission at radio frequencies and that the latter does not 
overshine the former in the optical. For this test, we make use of equations (\ref{nuic}), (\ref{fpjet}), 
and (\ref{RS}) for the evolution of the spectral characteristics, noting that the cooling frequency is 
the same for both shocks, as the internal energy density, bulk Lorentz factor, and comoving frame fireball
age have the same values behind both shocks.

 From equation (\ref{spek}), the reverse shock optical emission at $t > t_r$, when $\nuiRS$ is below the 
radio domain, is
\begin{equation}
  F_o^{(R)} (t) = F_r(t) \left( \frac{\nu_r}{\nu_o} \right)^{(p-1)/2} 
                 \min \left\{ 1, \left( \frac{\nu_c(t)}{\nu_o} \right)^{1/2} \right\} 
                 \quad {\rm where} \quad F_r(t) = \Frtr \left(\frac{t}{t_r}\right)^{-\alpha_r} \;.
\end{equation}
The condition that the slower decaying reverse shock emission does not overshine that from the forward 
shock, $F_o^{(F)}(t) = \Fotb (t/t_b)^{-\alpha_o}$, until the last optical measurement ($t_{max}$), leads 
to an upper limit on the cooling frequency:
\begin{equation}
  \log \frac{\nuctb}{\nu_o} < 2 \log \frac{\Fotb}{\Frtr} + (p-1) \log \frac{\nu_o}{\nu_r}
             + \frac{4-3s}{8-2s} (1+e) \log \frac{t_{max}}{t_b} + 2 \alpha_r \log \frac{t_{max}}{t_r}
             + 2 \alpha_o \log \frac{t_b}{t_{max}} \;. 
\label{nucc}
\end{equation}

 As for the case of a structured outflow (\S\ref{structure}), the radio emission $F_r^{(R)}$ from the reverse 
shock is not overshined by that from the forward shock, $F_r^{(F)}$, if $F_r^{(R)} (t_+) > F_r^{(F)} (t_+)$, 
where $t_+$ is the time when $\nuiFS$ crosses the radio domain. This condition leads to the same lower 
limit for the forward shock injection frequency ($\nuiFStb$) as in equation (\ref{nuiO}), except that now
\begin{equation}
 \xi(e) \equiv \frac{p}{2} + \frac{2}{3-e} \left[ \frac{3-e(13-4s)}{3(4-s)}-\alpha_r \right] - \frac{1}{6} \;,
\label{xie}
\end{equation}
and the cooling frequency is that given in equation (\ref{nucc}). 
The lower limit on $\nuiFStb$ implies a lower limit on the forward shock injection frequency at the 
time $t_{min}$ of the first optical observation, $\nuiFS(t_{min}) = \nuiFStb (t_b/t_{min})^{(3-e)/2}$,
which must be redward of the optical domain, to explain the decay of the optical light-curve. This 
requirement provides a non-trivial test for the reverse--forward shock scenario, as there are cases 
where, even for the highest injection frequency allowed by equation (\ref{nui}), the peak flux of the 
forward shock emission at the when $\nuiFS$ reaches the radio domain is brighter than the observed radio 
flux (attributed to the reverse shock). The above cases that satisfy this constraint are:
\begin{enumerate}
 \item GRB 991208 
   \begin{itemize} \vspace*{-2mm}
    \item $s=0$, $\nuiFStb > 3\times 10^{12}$ Hz, $\nuctb < \nu_o$,
    \item $s=2$, $\nuiFStb > 3\times 10^{11}$ Hz, $\nuctb < 3\times 10^{14}$ Hz,
   \end{itemize}
 \item GRB 991216, for $\beta_o = 0.58$
   \begin{itemize} \vspace*{-2mm}
    \item $s=0$, $\nuiFStb > 10^{13}$ Hz, $\nuctb > \nu_o$,
    \item $s=2$, $\nuiFStb > 6\times 10^{12}$ Hz, $\nuctb > \nu_o$,
   \end{itemize}
  \vspace*{-2mm} \hspace*{8mm}  
    while for $\beta_o = 0.87$, $s=0$, $\nuiFStb > 2\times 10^{12}$ Hz, $\nuctb < \nu_o$,
  \vspace*{1mm}
 \item GRB 000926, $s=0$, $\nuiFStb > 4\times 10^{12}$ Hz, $\nuctb > \nu_o$,
  \vspace*{1mm}
 \item GRB 010222,
   \begin{itemize} \vspace*{-2mm}
    \item $s=0$, $\nuiFStb > 7\times 10^{12}$ Hz, $\nuctb < 6\times 10^{13}$ Hz,
    \item $s=2$, $\nuiFStb > 10^{13}$ Hz, $\nuctb < 50$ GHz.
   \end{itemize}
\end{enumerate}

 Finally, the reverse--forward shock scenario must satisfy two other constraints: the observed
optical flux at the break time, $\Fotb$, and the passage through the radio domain of the reverse
shock frequency $\nuiRS$ at time $t_r$ when the radio light-curve begins to fall-off.
The former determines the peak flux of the forward shock emission:
\begin{equation}
 F_p^{(F)} (t_b) = \Fotb \left[ \frac{\nu_o}{\nuiFStb} \right]^{(p-1)/2} 
                   \max \left\{1, \left( \frac{\nu_o}{\nuctb} \right)^{1/2} \right\} \;.
\label{fpF}
\end{equation}
The latter constrains the fireball Lorentz factor, as following. If the microphysical parameters
$\epsi$ and $\epsB$ have the same values behind both shocks, then the ratio of the reverse shock 
to the forward shock injection frequencies is 
\begin{equation}
  \frac{\nuiRS}{\nuiFS} = \left( \frac{\Gamma'-1}{\Gamma} \right)^2  \quad {\rm where} \quad
  \Gamma' = 1 + \frac{(\Gamma_i - \Gamma)^2}{2 \Gamma_i \Gamma} = \frac{5+e-s}{2[(4-s)(e+1)]^{1/2}} \;,
\label{gmprime}
\end{equation}
is the Lorentz factor of the shocked ejecta in the frame of the incoming, undecelerated ejecta.
Note that for $e=0$, $\Gamma' = 1.25$ for $s=0$ and $\Gamma' = 1.06$ for $s=2$, therefore the reverse 
shock crossing the ejecta is semi-relativistic. Taking into account the evolution of the reverse shock
injection frequency (\eq [\ref{RS}]) and equation (\ref{gmprime}), the condition $\nuiRS(t_r) = \nu_r$
is equivalent to  
\begin{equation}
  \Gamma (t_b) = \frac{(5+e-s)^2}{4(4-s)(e+1)} \left( \frac{t_b}{t_r} \right)^{\frac{6-s-e(2-s)}{16-4s}} 
                 \left[ \frac{\nuiFStb}{\nu_r} \right]^{1/2} \;. 
\label{Gmtb}
\end{equation}

 With the aid of equations (\ref{nuiO}), (\ref{nucc}), (\ref{xie}), (\ref{fpF}), and (\ref{Gmtb}), one 
can determine $\nuiFStb$, $\nuctb$, $F_p^{(F)} (t_b)$, and $\Gamma (t_b)$ from afterglow observations,
which can be turned into four constraints for the four basic afterglow parameters $\E$, $n$ (or $A_*$ 
for a wind), $\epsB$, and $\epsi$, after using equations (\ref{nui}), (\ref{nuc0}), (\ref{nuc2}), 
(\ref{fp0}), (\ref{fp2}), and the expressions for the fireball Lorentz factor (\eg Panaitescu \& Kumar 2000): 
\begin{equation}
 \Gamma (t) = 8\, (z+1)^{3/8} \left(\frac{\E}{10^{53}}\right)^{1/8} \left(\frac{n}{1\cm3}\right)^{-1/8}\,
                     t_d^{-3/8} \;\; {\rm for} \;\; s=0, \quad
 \Gamma (t) = 15\,(z+1)^{1/4} \left(\frac{\E}{10^{53}}\right)^{1/4} A_*^{-1/4}\, t_d^{-1/4} \;\; 
                {\rm for} \;\; s=2 \; .
\label{Gamma}
\end{equation}
From here, the allowed ranges of each parameter can be used as a test of the reverse--forward shock 
scenario by requiring that $\epsi$ and $\epsB$ are below unity, and that $\E$ and $n$ (or $A_*$) 
have plausible values. For the cases outlined above, involving a homogeneous medium, we find the 
following constraints on the afterglow parameters
\begin{enumerate}
 \item GRB 991208, $\E < 10^{51}$ ergs/sr, $n \in (0.01, 10) \cm3$, $\epsi \in (0.05,0.2)$, $\epsB  > 0.04$, 
 \item GRB 991216, $\E < 2\times 10^{52}$ ergs/sr, $n < 30 \cm3$, $\epsi \in (0.05,0.2)$, $\epsB > 0.02$,
 \item GRB 000926, $\E < 6\times 10^{51}$ ergs/sr, $n < 50 \cm3$, $\epsi \in (0.06,0.3)$, $\epsB > 0.05$,
 \item GRB 010222, $\E < 3\times 10^{50}$ ergs/sr, $n \in (0.05,3) \cm3$, $\epsi \in (0.03,0.07)$, $\epsB > 0.1$.
\end{enumerate}

 Compared to the afterglow parameters found by us (Panaitescu \& Kumar 2002) through fitting
the radio, optical and $X$-ray light-curves of these afterglows with the forward shock model 
(\ie without any contribution from the reverse shock), the above kinetic energies per solid
angle for the reverse--forward shock model are slightly lower, the magnetic field parameter
is somewhat larger, while the circumburst densities and electron energy parameter are similar. 
The largest discrepancy occurs for the afterglow 010222, for which radio data were not available
at the time when our afterglow fits were done.

For a wind-like medium, the required kinetic energies are below $10^{50}$ ergs/sr, the microphysical 
parameters are slightly larger, and the wind density corresponds to $A_* \in (0.1, 1)$. We note that 
kinetic energies under $10^{50}$ ergs/sr require an extremely efficient GRB mechanism to yield the 
observed $\gamma$-ray outputs, ranging from $10^{52}$ to $5 \times 10^{52}$ ergs/sr (Bloom, Frail \& 
Sari 2001) for the above four afterglows, thus a wind-like medium is somewhat less favoured.

\subsection{Non-Relativistic Outflows}

 The scenarios considered so far involved relativistic fireballs. For either a spherical outflow 
or a jet with sharp edges, the afterglow light-curves asymptotically reach the same decay as
the outflow becomes non-relativistic because the lateral spreading of the jet diminishes and
the entire emitting surface is visible in both cases. As shown by Vietri \& Waxman (2000), for 
a jet, the transition to the non-relativistic regime mitigates the light-curve decay, while for 
a fireball it leads to a steeper decay. This may suggest that a slow radio decay following a 
fast optical fall-off may be explained by a jet becoming non-relativistic (Frail \etal 2004). 

 In the above scenario, the optical light-curve is $F_o \propto t^{-p}$ (relativistic jet). 
From the basic equations for dynamics and evolution of spectral parameters, the radio flux
from a non-relativistic fireball is found to be
\begin{equation}
  F_{\nu_i < \nu < \nu_c} \propto t^{-3(5p-7)/10}  \;\; {\rm for\;\; s=0} \;;  \quad  
  F_{\nu_i < \nu < \nu_c} \propto t^{-(7p-5)/6}   \;\; {\rm for\;\; s=2} \;.
\label{fnr}
\end{equation}
For the electron distribution index implied by the optical light-curve index $\alpha_o$ (Table 2), 
the resulting radio decay is significantly steeper than observed for the afterglows 000301 and
000926, for either type of medium. For the afterglows 991208, 991216, and 010222, a homogeneous
medium leads to radio decays shallower than observed, while a wind medium yields radio fall-offs
that are too steep. Since the non-relativistic transition flattens the decay of the emission from
a jet, it can be argued that a semi-relativistic motion leads to a milder flattening than expected
for the non-relativistic regime, becoming thus compatible with the afterglows 991208, 991216, and 
010222, for which the expected asymptotic radio decay is too shallow in the case of a homogeneous
medium.

 A severe problem with the scenario of a relativistic optical and semi-relativistic radio 
afterglow is that it requires the shallow radio decay to be seen after the steeper optical
fall-off. Therefore this scenario cannot be at work in the afterglows 991216 and 010222, whose
radio and optical observations overlap in time over one and almost two decades, respectively. 
For the afterglows 991208, 000301, and 000926, the overlap is short-lived, still it is rather 
unlikely that, during the short time interval from the last optical measurements and untill 
the onset of the radio decay, a relativistic jet decelerating as $\Gamma \propto t^{-1/2}$ can 
slow down to a semi-relativistic motion.

 We note that a scenario where the fireball becomes non-relativistic before the onset of the 
optical fall-off cannot explain the observed index difference $\Delta \alpha_{or}$.
Above the cooling frequency, the light-curve for a non-relativistic afterglow is
\begin{equation}
  F_{\nu_c < \nu} \propto t^{-3(p-2)/4}  \;\; {\rm for\;\; s=0} \;;  \quad  
  F_{\nu_c < \nu} \propto t^{-(7p-8)/6}  \;\; {\rm for\;\; s=2} \;.
\end{equation}
Using equation (\ref{fnr}), it follows that, when the cooling frequency is between the radio and
optical domains, $\Delta \alpha_{or} = 0.1$ for a homogeneous medium and $\Delta \alpha_{or} = -0.5$
for a wind, neither of which is compatible with the observations of the five anomalous afterglows.

 As the non-relativistic motion of a uniform GRB remnant seems unable to accommodate the five 
anomalous afterglows, we turn our attention to a structured outflow. We consider again the dual
outflow scenario, which allows a maximal decoupling of the radio emission, arising from the
non-relativistic, uniform outflow envelope, from the optical emission produced by a relativistic,
uniform core. Equation (\ref{Gamma}) shows that, for the envelope to become non-relativistic at 
$t_{nr} \sim 10$ days after the burst, the external density must be $n \sim 10^5 (\E_R/10^{53})\,\cm3$ 
or the GRB progenitor must have a wind with $A_* \sim 10^4 (\E_R/10^{53})$ (\ie much larger than for 
any known type of massive star), where $\E_R$ is the kinetic energy per solid angle in the radio envelope. 

 Substituting in equations (\ref{nui}), (\ref{nua0a}), and (\ref{nua2a}), such high densities 
lead to an optically thick afterglow at the peak of the synchrotron spectrum at $t = t_{nr}$, 
therefore the self-absorption frequency is given by $\nu_a = \nu_i \tau_i^{2/(p+4)}$, instead of
equations (\ref{nua0a})--(\ref{nua2a}), where $\tau_i = (5 e \Sigma)/(B \gamma_i^5)$ is the optical 
thickness to self-absorption at the injection frequency, $\Sigma \propto n r$ being the electron 
column density of the swept-up circumburst medium, and $e$ the electron charge.
For $p \sim 2$ and a homogeneous medium with the high density required by the early onset of the 
non-relativistic regime, the resulting self-absorption frequency is 
\begin{equation}
 \nu_a (t_{nr}) \sim 600 \, \left( \frac{\E_R}{10^{53}} \right)^{2/3} 
                            \left( \frac{\epsi}{0.03} \right)^{1/3} 
                            \left( \frac{\epsB}{10^{-3}} \right)^{1/3} \; {\rm GHz} \;.
\label{nua}
\end{equation}

 If the radio domain were self-absorbed at $t > t_{nr}$, the radio emission would rise as 
$F_r \propto t^{1.1}$ (for $p=2$; Frail, Waxman \& Kulkarni 2000) during the non-relativistic phase. 
Therefore the observed radio fall-off implies that the radio emission from the envelope is optically 
thin at $t = t_{nr}$. Then, equation (\ref{nua}) leads to $\E_R < 2\times 10^{50} (\epsi/0.03)^{-1/2} 
(\epsB/10^{-3})^{-1/2}$ erg/sr. 
For such a low energy per solid angle in the envelope, the condition for non-relativistic
motion at $t = 10$ days becomes $n \sim 300\, (\epsi/0.03)^{-1/2} (\epsB/10^{-3})^{-1/2} \, \cm3$.
Making use of equation (\ref{Gamma}), for this circumburst density, the optical core must have an 
energy per solid angle $\E_O \simg 10^{54} (\epsi/0.03)^{1/2} (\epsB/10^{-3})^{1/2}$ erg/sr for
the core to remain relativistic untill at least 10 days and accommodate the observed steep optical 
decay. Assuming the same microphysical parameters in the entire outflow, that $F_p \propto \E$ 
(\eq [\ref{fp0}]) implies that the peak flux of the core emission is much larger than that of the 
envelope emission. As it softens, the core emission should become brighter at radio frequencies 
than the envelope, therefore this scenario cannot decouple the radio and optical emissions and 
cannot account for the five anomalous afterglows.

\section{Conclusions}

 As shown in Figure 1, the power-law decays of the radio and optical 
emissions of the GRB afterglows 970508, 980703, 000418, and 021004 are similar, as is expected in
the simplest form of the relativistic fireball model (adiabatic dynamics, uniform outflow, no energy 
injection, constant microphysical parameters). The shallow decay of the radio emission of the afterglow 
010222 is only marginally compatible with that from a spreading jet but requires some extreme parameters 
(\S\ref{jet2}). However, the shallow radio decay and the steep optical fall-off of the afterglows 
991208, 991216, 000301, and 000926, cannot be explained within the simplest afterglow model (see also 
Frail \etal 2004).  The reason is that, for this model to explain the observations, the injection 
frequency $\nu_i$ should be between the radio and optical domains until about 100 days, which, as 
shown in \S\ref{jet1} and \S\ref{fballwind}, leads to a radio emission much brighter than observed. 

 We are thus forced to consider scenarios that decouple the radio and optical decay indices. 
Assuming that the radio and optical emissions arise from the same population of electrons energized
by the forward shock, we have investigated two scenarios involving microphysical parameters that vary 
in time. Alternatively, the low and high frequency emissions may arise in different parts of the 
fireball if the outflow has an angular structure or if there is a long-lived reverse shock. 

 In the former category, we considered the case where there is no spectral break between the radio
and optical domains, requiring a variable slope of the synchrotron spectrum, \ie a time-varying slope 
of the electron power-law energy distribution $dN/d\gamma \propto \gamma^{-p}$. The evolution of $p$
can be determined (\S\ref{varp}) by requiring that the resulting optical light-curve is a power-law in 
time, as the longest time coverage with the least measurement errors is achieved in this domain.
For the five afterglows with shallow radio decays, we obtain that $p$ should increase in time, which 
implies that the optical spectrum should soften in time, the radio light-curves decay should steepen 
($d^2 F_r/dt^2 < 0$), while $X$-ray light-curves should flatten ($d^2 F_x/dt^2 > 0$). With the exception 
of the afterglow 000301, the resulting radio and $X$-ray light-curve decays and the optical spectral 
slope are in conflict with the observations.

 Also in the first category, we investigated a scenario where the cooling frequency $\nu_c$ is between
the radio and optical domains. In this case the optical and $X$-ray decays should to be identical,
as for the afterglows 000926 and 010222, but inconsistent with the observations for 991216.
To explain the shallowness of the radio decays and the steepness of the optical light-curves, 
$\nu_c$ must decrease faster than usually expected ($\nu_c \propto t^{-1/2}$ for a homogeneous medium), 
which requires an increasing magnetic field parameter $\epsB$. The evolution of $\epsB$ and 
that of the parameter $\epsi$ for minimal electron energy were determined from the observed radio and 
optical decay indices (\S\ref{vareps}). We found that, if there is no energy injection, for any geometry 
of the ejecta (spherical or collimated) and any radiative regime, $\epsi$ must decrease and $\epsB$ must 
increase with time\footnote{ 
  By modeling the broadband emission of four GRB afterglows with evolving $\epsB$, Yost \etal (2003) 
  have found that good fits can be obtained for $\epsB$ decreasing slower than $t^{-0.4}$ or increasing 
  slower than $t^{3/4}$}. 
For both a homogeneous or a wind-like medium, the increase of $\epsB$ is so strong that the magnetic 
field $B$ is either constant or increases in time, which is not consistent with a field arising in a
turbulent, decelerating flow, but could be compatible with a GRB occurring in a pulsar bubble 
(Konigl \& Granot 2002), where the fireball interacts with a magnetized medium, provided that a
supernova occurred years before the burst. If there is energy injection, 
solutions with constant $\epsB$ are possible; however, this scenario leads to evolutions of the peak 
flux $F_p$ and injection frequency $\nu_i$ which are in contradiction with the behaviors inferred by 
Galama \etal (2001) for the afterglow 991208 from its radio observations. A magnetic field strength 
that decreases in time can obtained if the density of the circumburst medium increases with radius.  
In this case, the resulting evolutions of $F_p$ and $\nu_i$ are more consistent with the observations 
of 991208 if the outflow is collimated. 

 We did not considered a scenario where all microphysical parameters ($\epsi$, $\epsB$, $p$)
evolve in time, such that the difference $\alpha_o - \alpha_r$ observed between the radio and
optical decay indices is due both to a steepening electron distribution and a cooling frequency
decreasing faster than usually. The increase of the electron index $p$ would reduce the required 
increase of $\epsB$, though a significant evolution of the former would probably lead to $X$-ray
light-curve decays that are too steep compared with those observed for the afterglows 000926
and 010222.

 In the second type of scenarios, we have considered a dual outflow endowed with a uniform core of 
higher energy per solid angle, emitting the optical afterglow, surrounded by a uniform envelope of 
a lower energy density, releasing the radio afterglow. This scenario allows a better decoupling
of the radio and optical emissions than for an outflow with a smooth angular structure, and finds 
support in the steepenings ("breaks") observed in the decay of the optical afterglows 991216, 000301, 
000926, and 010222, at around 1 day after the burst. Such breaks are expected when the boundary of 
the optical core becomes visible to the observer. 
 Another possibility is that the low frequency afterglow emission arises in the semi-relativistic 
reverse shock occurring when there is a continuous inflow of ejecta into the fireball, while the 
higher frequency emission is (as usually) attributed to the forward shock. To explain the observed 
steepening of the optical decay, we have maintained the premise of a structured outflow, but we have 
ignored the radiation emitted by the envelope. 

 Within the framework of these two scenarios, we required that the softening of the emission 
from the optically radiating region does not make its radio contribution overshine the low frequency 
emission from the radio emitting region. This requirement sets a lower bound (\eq [\ref{nuiO}]) 
on the injection frequency $\nu_i (t_b)$ for the optical component at the time $t_b$ ($\sim 1$ day) 
when the optical light-curve break is observed. At the same time, we required that the shallower
decaying emission from the radio component does not yield an optical contribution exceeding that
arising from the optical region. This condition leads to an upper limit on the cooling frequency
$\nu_c (t_b)$ for the radio outflow (\eq [\ref{nucc}]).

 For the anisotropic outflow scenario (\S\ref{structure}), a stringent test is provided by the lower 
limit (\eq [\ref{nuiR}]) on the injection frequency of the radio emitting envelope which can be derived 
from that of the optical core. Since the onset of the radio decay is due to the passage of the former 
injection frequency through the radio domain, the lower limit on it should be below 10 GHz. For the 
four afterglows (991208, 991216, 000926, 010222) with slowly decaying radio emission and optical 
breaks compatible with a structured outflow, this condition is at best marginally satisfied. In those 
marginal cases, total electron energies exceeding equipartition with protons and very dense external 
media, leading to a self-absorbed radio emission and semi-relativistic fireball. Therefore, the 
structured outflow scenario cannot account for all the radio and optical properties of the above four 
afterglows. 

 In the reverse--forward shock scenario (\S\ref{reverse}), we found that the sets of parameters 
describing the energy and mass injection which are self-consistent and which can explain the magnitude 
of the break observed in optical light-curves, correspond to a modest increase in the fireball energy, 
by less than a factor 2. To explain the shallowness of the radio decay, the incoming ejecta must have 
an energy distribution which increases rapidly with decreasing Lorentz factor: $d\E_i/d\Gamma_i \propto 
\Gamma_i^{-(2.5 \div 4)}$ for a homogeneous medium and $\propto \Gamma_i^{-(4 \div 5)}$ for a wind-like 
medium. An important constraint is provided by that the afterglow optical light-curves decay from the 
time $t_{min}$ of the first measurement, which requires that the lower bound on the forward shock 
injection frequency is redward of the optical domain at $t_{min}$. This lower limit, the upper limit on the 
cooling frequency (see above), the observed optical flux, and the condition that the reverse shock 
injection frequency crosses the radio domain when the radio decay starts, provide four constraints for 
the fireball kinetic energy per solid angle ($\E$), external particle density ($n$), and microphysical
parameters $\epsi$, and $\epsB$. We found that, to explain the properties of the radio and optical
light-curves of the afterglows 991208, 991216, 000926 and 010222, the reverse--forward shock scenario
requires total electron energy and magnetic field parameters which are close to equipartition, circumburst 
particle densities between $10^{-3} \cm3$ and $10\cm3$, and low fireball kinetic energies (upper limits 
ranging from $3\times 10^{50}$ to $3\times 10^{52}$ ergs/sr), which require a very efficient GRB mechanism. 
These parameters are roughly consistent with those inferred through modeling of the afterglow broadband 
emission with the external shock only (Panaitescu \& Kumar 2002). 
Overall, the reverse--forward shock scenario is the most promising of all scenarios discussed here in 
accommodating the shallow radio decay observed for the five anomalous GRB afterglows.

 A possibility not discussed so far is that the anomalous radio/optical afterglows are due to a new 
type of spectral break located between the radio and optical domains. Such a feature has been used 
to explain the broadband emission of the afterglows 991208 and 000301 (Li \& Chevalier 2001, Panaitescu 
2001). The passage of this break frequency ($\nu_*$) through the optical domain yields a softening of 
the optical spectral slope simultaneous to the light-curve break, as seems to have been the case for 
the afterglow 000301. To determine the evolution of $\nu_*$ requires a measurement of the optical
spectral slope $\nu_*$ is redward of the optical domain. Such measurements exist only for the 
afterglows 991208 and 991216, for which we find that $\nu_* \propto t^{-2}$ for a homogeneous medium 
and $\nu_* \propto t^{-1}$ for a wind-like medium, assuming that the ejecta are spherical. These 
scalings correspond to a break in the electron distribution at energy $\gamma_* m_e c^2$ satisfying 
$\gamma_* \propto \Gamma^{3.5}$ and $\gamma_* \propto \Gamma^{2/3}$, respectively, where $\Gamma$ 
is the fireball Lorentz factor. While the former scaling has no obvious interpretation, the latter 
is not far from the $\gamma_* \propto \Gamma$ expected for $p < 2$, if the total electron energy is 
assumed to be a constant fraction of the post-shock energy.

\vspace*{5mm} \noindent
{\bf Acknowledgments.}
 We thank Shri Kulkarni and Dale Frail for interesting and stimulating discussions about GRB afterglows.

\vspace*{1cm}
\begin{center}
 \parbox[t]{15cm}{
  {\bf TABLE 1.} {\sc Definitions of quantities pertaining to afterglow dynamics and radiation}
  }  \\ [4ex]
\begin{tabular}{ll}
 \hline \hline
 \rule[-2mm]{0mm}{6mm} Notation  & Definition \\ \hline

 \rule[-2mm]{0mm}{8mm} $\nu_a$   & synchrotron self-absorption frequency \\
 \rule[-2mm]{0mm}{5mm} $\nu_i$   & injection frequency (synchrotron characteristic frequency for the typical
                                     electron energy)  \\
 \rule[-2mm]{0mm}{5mm} $\nu_c$   & cooling frequency (synchrotron characteristic frequency for electrons
                                    whose radiative cooling timescale equals the fireball age) \\
 \rule[-2mm]{0mm}{5mm} $F_p$     & flux at the peak of the synchrotron spectrum $F_\nu$  \\

 \rule[-2mm]{0mm}{5mm} $\alpha$  & index of the afterglow power-law light-curve $F_\nu \propto t^{-\alpha}$ \\
 \rule[-2mm]{0mm}{5mm} $\delta \alpha_o$ & increase $\alpha_o (t > t_b) - \alpha_o (t < t_b)$ of the optical 
                                   decay index at $t_b$ \\ 
 \rule[-2mm]{0mm}{5mm} $\beta$   & slope of the afterglow power-law spectrum $F_\nu \propto \nu^{-\beta}$ \\

 \rule[-2mm]{0mm}{5mm} $\E$      & fireball kinetic energy per solid angle, $^{(O)}$ for the optical core,
                                   $^{(R)}$ for the radio envelope \\
 \rule[-2mm]{0mm}{5mm} $p$       & index of power-law electron energy distribution, 
                                     $dN_e/d\gamma \propto \gamma^{-p}$ \\
 \rule[-2mm]{0mm}{5mm} $\epsi$   & parameter for the minimum electron energy. 
                                   For $p > 2$, the electron total energy is a fraction $\frac{p-1}{p-2} \epsi$ 
                                   of the post-shock energy  \\
 \rule[-2mm]{0mm}{5mm} $\epsB$   & fraction of post-shock energy imparted to the magnetic field \\
 \rule[-2mm]{0mm}{5mm} $s$       & index of circumburst medium particle density, $n(r) \propto r^{-s}$ \\

 \rule[-2mm]{0mm}{5mm} $\nu_r, \nu_o$ & typical frequency for the radio and optical domains, respectively \\
 \rule[-2mm]{0mm}{5mm} $^{(R)}$  & superscript for the radio emitting outflow envelope or reverse shock  \\
 \rule[-2mm]{0mm}{5mm} $^{(O)}$  & superscript for the optically emitting outflow core \\
 \rule[-2mm]{0mm}{5mm} $^{(F)}$  & superscript for the forward shock \\

 \rule[-2mm]{0mm}{5mm} $t_{min}, t_{max}$ & epoch of first and last optical measurement, respectively \\
 \rule[-2mm]{0mm}{5mm} $t_b$     & time when the optical light-curve decay steepens \\
 \rule[-2mm]{0mm}{5mm} $\Fotb$   & optical flux at the break-time \\
 \rule[-2mm]{0mm}{5mm} $t_r$     & time when the radio light-curve begins to fall-off \\
 \rule[-2mm]{0mm}{5mm} $\Frtr$   & radio flux at onset of radio decay \\
 \hline \hline
\end{tabular}
\end{center}

\clearpage

\begin{table}
\vspace*{1cm}
\begin{center}
 \parbox[t]{13cm}{
    {\bf TABLE 2.} {\sc Temporal scaling of the synchrotron flux from an adiabatic fireball at an observing
            frequency $\nu$ for the three orderings of the synchrotron break frequencies $\nu_a$, 
             $\nu_i$, and $\nu_c$ that may occur at 1--100 days after the burst} 
   } \\ [4ex]
\begin{tabular}{c|ccc|ccc|ccc}
 \hline \hline
 \rule[-2.5mm]{0mm}{8mm}
            & \multicolumn{3}{c|}{\rm fireball, $n \propto r^0$} & 
              \multicolumn{3}{c|}{\rm fireball, $n \propto r^{-2}$} &
              \multicolumn{3}{c}{\rm jet} \\
 \rule[-2mm]{0mm}{4mm}
 {\rm case:} & (1) & (2) & (3) & (1) & (2) & (3) & (1) & (2) & (3)  \\ \hline
 \rule[-2.5mm]{0mm}{8mm} 
    $\nu_a<\nu_c<\nu_i$ & $t^{1}$   & $t^{1/6}$ & $t^{-1/4}$      & $t^{2}$ & $t^{-2/3}$ & $t^{-1/4}$ 
              & $t^{1}$ & $t^{-1}$  & $t^{-1}$   \\ \hline
 \rule[-2.5mm]{0mm}{8mm} 
    $\nu_a<\nu_i<\nu_c$ & $t^{1/2}$ & $t^{1/2}$ & $t^{-(3p-3)/4}$ & $t^{1}$ & $t^{0}$    & $t^{-(3p-1)/4}$ 
              & $t^{0}$ &$t^{-1/3}$ & $t^{-p}$   \\ \hline
 \rule[-2.5mm]{0mm}{8mm} 
    $\nu_i<\nu_a<\nu_c$ & $t^{1/2}$ & $t^{5/4}$ & $t^{-(3p-3)/4}$ & $t^{1}$ & $t^{7/4}$  & $t^{-(3p-1)/4}$ 
              & $t^{0}$ & $t^{1}$   & $t^{-p}$   \\ \hline 
  \hline
\end{tabular}
\end{center}

\hspace*{25mm} \parbox[t]{13cm}{
 (1): $\nu$ below the lowest spectral break \\
 (2): $\nu$ above the lowest break but below the mid break, \\ 
 (3): $\nu$ above the mid spectral break but below the highest
}
\end{table}

\begin{table}
\vspace*{1cm}
\begin{center}
 \parbox[t]{13cm}{
   {\bf TABLE 3.} {\sc Parameters of radio, optical, and $X$-ray light-curves (see Table 1 for definitions). 
                 Uncertainties of decay indices $\alpha$ are given in Figure 1. 
                 Times in days, fluxes in} mJy.  
   } \\ [4ex]
\begin{tabular}{c|ccc|cc|ccc|cc}
 \hline \hline
 \rule[-2.5mm]{0mm}{8mm}
   GRB  & $\alpha_r$ & $\alpha_o\dagger$ & $\alpha_x$ & $t_o$ & $\beta_o\ddagger$ & $t_b$ & $F_o(t_b)$ & 
                 $\delta \alpha_o$ & $t_r$ & $F_r(t_r)$ \\ \hline 
 \rule[-2.5mm]{0mm}{8mm} 
  991208 &  0.97  &  1.78  & .....  &  4  & $1.05\pm0.09^a$ & $<2$ & $>0.1$ &     .....     &  10  &  1   \\
 \rule[-2.5mm]{0mm}{8mm} 
  991216 &  0.69  &  1.59  &  1.82  &  2  & $0.58/0.87^b$   &  1   &  0.2   & $0.37\pm0.06$ & $<2$ & $>1$ \\
 \rule[-2.5mm]{0mm}{8mm} 
  000301 &  1.26  &  2.68  & .....  &  3  & $0.57\pm0.02^c$ &  5   & 0.015  & $2.03\pm0.08$ &  30  & 0.3  \\
 \rule[-2.5mm]{0mm}{8mm} 
  000926 &  0.68  &  2.30  &  2.19  &  1  & $1.00\pm0.18^d$ & 1.5  &  0.03  & $0.71\pm0.06$ &  20  & 0.4  \\
 \rule[-2.5mm]{0mm}{8mm} 
  010222 &  0.43  &  1.27  &  1.39  & 0.2 & $0.78\pm0.02^e$ & 0.6  &  0.06  & $0.60\pm0.06$ &   1   & 0.2  \\
 \hline \hline
\end{tabular}
\end{center}

\hspace*{25mm} \parbox[t]{13cm}{ 
 $\dagger$ optical light-curve decay index measured after the break, \ie at $t > t_b$ \\
 $\ddagger$ slope of optical spectrum determined at epoch $t_o$ \\
 $^a$ from Castro-Tirado \etal (2001) \\
 $^b$ $0.58\pm0.08$ for Galactic reddening of $E(B-V)=0.63$, $0.87\pm0.08$ for $E(B-V)=0.46$ \\
           (Garnavich \etal 2000, Halpern \etal 2000) \\
 $^c$ after correction for host extinction of $A_V=0.14\pm0.01$ (Jensen \etal 2001) \\
 $^d$ after correction for host extinction of $A_V=0.18\pm0.06$ (Fynbo \etal 2001) \\
 $^e$ after correction for host extinction of $A_V \sim 0.1$ (Lee \etal 2001, Galama \etal 2003) 
}

\end{table}

\begin{figure}
\vspace*{1cm}
\centerline{\psfig{figure=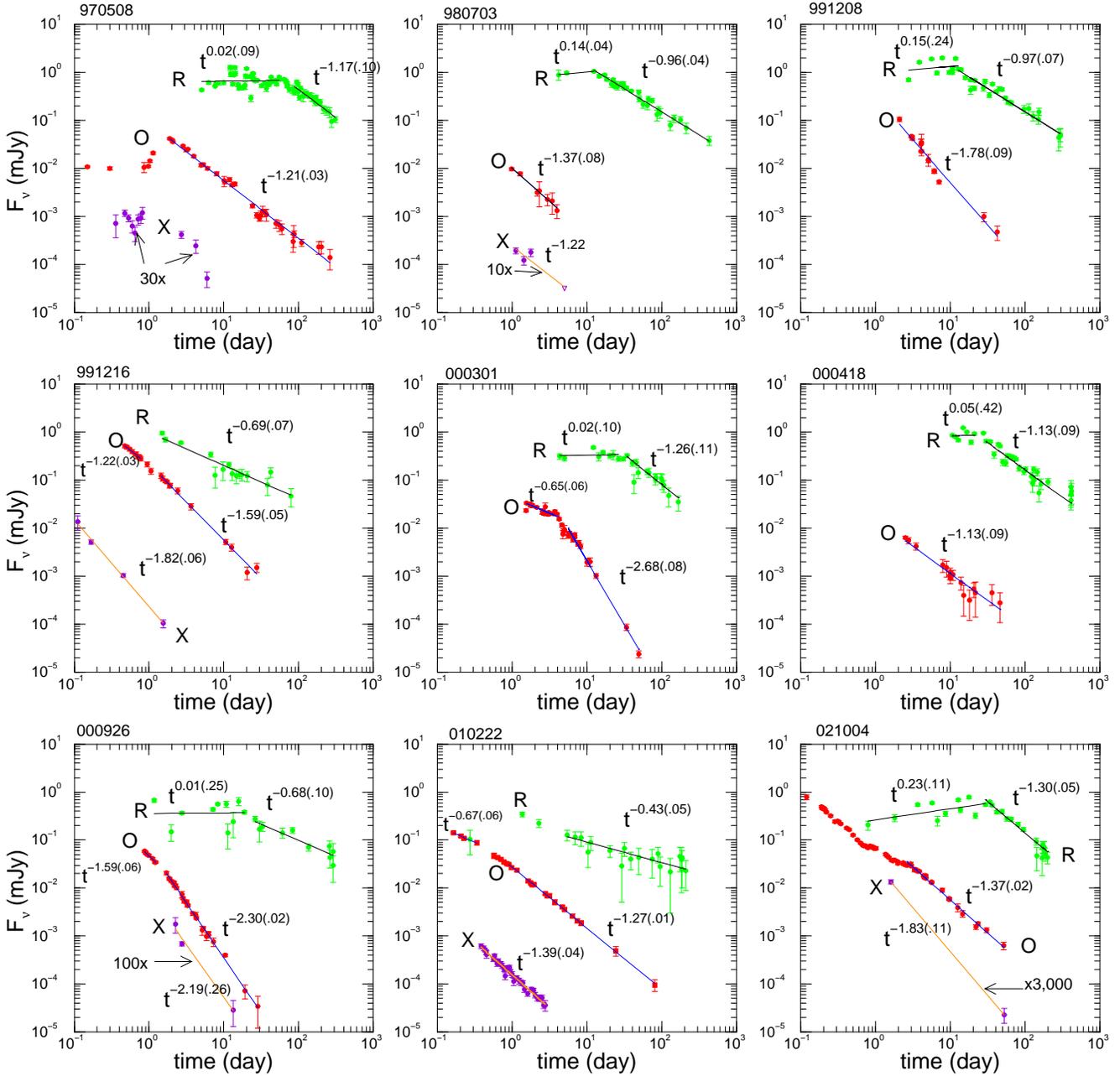}}
\vspace*{1cm}
\caption{\small Radio (8 GHz), Optical ($5 \times 10^{14}$ Hz), and X-ray (5 keV) light-curves 
  of all GRB afterglows with well-monitored light-curves in the two lower frequency domains. 
  Power-law fits are indicated for each frequency range, errors are given in parentheses.
  If the last two optical measurements for 991208 are excluded, then the optical decay is
  $t^{-2.40\pm0.16}$. Some $X$-ray light-curves have been shifted upward by the indicated factors. 
  Note that power-law decay slope of the radio emission of 991208, 991216, 000301, 000926, and 
  010222 differs from that of the corresponding optical fall-off slope by more than 1/4, 
  contrary to what is expected in the simplest fireball model at late times, when the injection 
  frequency should be redward of the radio domain.}
\end{figure}

\end{document}